\def\hst{{\sl HST}}
\def\nicm{{\sl NICMOS}}
\def\spit{{\sl Spitzer}}
\def\irac{{\sl IRAC}}
\def\siri{{\sl SIRIUS}}
\def\2mass{{\sl 2MASS}}
\def\chan{{\sl Chandra}}
\def\nict  {{\sl NIC3}}
\def\nics  {{\sl NIC2}}

\documentclass[12pt,preprint]{aastex}
\usepackage{pdfsync}
\usepackage{epsf}
\usepackage{color}
\usepackage{graphicx}
\usepackage{graphics}
\usepackage{epsfig}
\usepackage{longtable}

\begin{document}
\title{A Multiwavelength Study of Evolved Massive Stars in the Galactic Center}
\author{H. Dong$^{1,2}$, Q. D. Wang$^1$, M. R. Morris$^3$}

\affil{$^1$ Department of Astronomy, University of Massachusetts,
Amherst, MA, 01003}\affil{$^2$National Optical Astronomy Observatory,
Tucson, AZ, 85719}\affil{$^3$
Department of Physics and Astronomy, University of California, Los
Angeles, 90095}\affil{E-mail: hdong@noao.edu, wqd@astro.umass.edu}

\begin{abstract}
The central region of the Milky Way Galaxy provides a unique laboratory
for a systematic, spatially-resolved population study of evolved massive stars
of various types in a relatively high metallicity environment. 
We have conducted a multi-wavelength data analysis of 180 such stars or
candidates, most of which were drawn from a recent large-scale
\hst /\nicm\ narrow-band Paschen-$\alpha$ survey, plus additional 14 
Wolf-Rayet stars identified in earlier ground-based spectroscopic observations 
of the same field. The multi-wavelength data include broad-band infrared (IR)
photometry measurements 
from \hst /\nics, \siri, \2mass, \spit/\irac, and X-ray observations
from \chan. We correct for extinctions toward individual stars, improve the
Paschen-$\alpha$ line equivalent width measurements, quantify the
substantial mid-IR dust emission associated with WC stars, and find
\textit{X}-ray
counterparts. In the process, we identify 10 foreground sources, some of
which may be nearby cataclysmic variables. 
The WN stars in the Arches and Central clusters show 
correlations between the Paschen-$\alpha$ equivalent width and the 
adjacent continuum emission. However, the
WN stars in the latter cluster are systematically dimmer than those in
the Arches cluster, presumably due to the different ages 
of the two clusters. 
In the equivalent width-magnitude plot, WNL stars, WC stars and OB
supergiants roughly fall into three distinct regions. We estimate that
the
dust mass associated with individual WC stars in the Quintuplet cluster can 
reach $10^{-5}  M_\odot$, or more than one order of magnitude
 larger than previous estimates. Thus WC stars could be a significant
 source of dust in the galaxies of the early universe.
Nearly half of the evolved massive stars in the Galactic center
are located outside the three known massive stellar clusters. 
The ionization of several compact H{\small \rm
  II} regions can be accounted for by their enclosed 
individual evolved massive stars, which thus likely formed in isolation
or in small groups.
\end{abstract}

\section{Introduction}~\label{s:introduction}
Massive stars play an important role in galaxy formation and evolution. They 
are a primary source of ionizing photons, radiation pressure, mechanical
energy, and chemical enrichment of the interstellar medium (ISM). When a 
massive star reaches its evolved stage (off the main sequence; 
MS hereafter), such feedback has its greatest impact, profoundly 
shaping the local and possibly even galactic environment.

Stellar evolution theories, though with significant
uncertainties, suggest that there are several
phases through which a very massive star with an initial mass $\gtrsim 25 M_\odot$
should evolve between leaving the MS
and exploding as a supernova. Depending on initial stellar mass and
metallicity, the sequence of evolutionary stages might be: O (MS)
$\rightarrow$ OB supergiant $\rightarrow$ LBV (luminous blue variable) 
or Ofpe/WN9~\citep[also known as WN10-11;][]{smi94,boh99}
$\rightarrow$ WN $\rightarrow$ WC $\rightarrow$
SNIb/SNIc~\citep{lan94,cro95,cro07}. WN and WC represent
 two main classes of Wolf Rayet  (WR) stars: the spectrum of a WN star is characterized
by strong emission lines of nitrogen (N${\small \rm III-V}$) relative
to He${\small \rm I-II}$,
whereas a WC star exhibits strong carbon lines (C${\small \rm III-IV}$).
These two classes are further divided into early (WN2-5: WNE and WC4-6: WCE) 
and late (WN7-9: WNL and WC7-9: WCL) types. In particular, the WCL and WNL types 
seem to be favored in a high metallicity 
environment~\citep{had05}. The emission lines are believed to arise from
optically-thin stellar winds, which can also produce a flat  
free-free continuum strong enough to flatten the stellar spectral shape
in the mid-IR~\citep{wri75,mau10a}. In addition, significant dust emission is often
found around WCL stars~\citep{cro03} and makes them very red in the mid-IR 
(e.g., \citealt{tut06}). The spectra of Ofpe/WN9 stars are similar to 
those of OB supergiants, but show strong He${\small \rm II}  (4696\AA)$ emission 
lines. Stars in the LBV phase experience sporadic,  
giant eruptive activities and are much rarer than WR stars. Thus the LBV 
phase should be very short. Much is yet to be learned to quantify the strong
feedback from these classes of evolved massive stars (EMSs) as well as their 
evolution and relative populations. 

Existing studies of EMSs are very much 
limited by small sample sizes.  Because of their short
lifetimes, EMSs are found abundant only in starburst 
regions. However, even in nearby nuclear starburst galaxies such as 
M83~\citep{had05}, such regions cannot be adequately resolved to identify 
individual EMSs, especially if they are in compact 
massive stellar clusters. Local Group galaxies are mostly in
relatively quiescent states; large samples of EMSs 
have to be collected from diverse regions having different star formation histories
and environments. Such samples are not suitable for a statistical
study of EMSs (e.g., to determine their evolutionary dependence on 
initial stellar mass and metallicity). For example, in the solar/super-solar
metallicity range, beyond the Milky Way, only M31 has been 
studied and just 48 WRs have been found in the whole galaxy~\citep{mas03}. 

The Galactic center (GC) of the Milky Way~\citep[$\sim$8.0 ${\rm kpc}$,][]{ghe08} provides an excellent laboratory
for a systematic study of EMSs with a apparent super-solar metallicity 
~\citep[$\sim 1.4-2\times$ solar;][]{mar07,mar08}. 
Within the central 200 pc radius, the GC accounts for about 10\% of
the ionizing photons~\citep{fig04} and $\sim$25\% of the entire WR
population in the Galaxy~\citep{mau10c}. About half of the ionizing radiation
from the GC is attributable to the Arches, Quintuplet and Central clusters~\citep{lan01}.
These three massive star clusters have well-determined ages 
(Arches: $\sim$ 2.5 Myr,~\citealt{fig02,naj04,mar08}, Quintuplet: 
$\sim$ 3.5 Myr,~\citealt{fig99a,lie12}, and Central: $\sim$ 6 Myr,~\citealt{gen03,pau06}). Each has a total
stellar mass of 
$\sim~10^4~M_{\odot}$. They contain numerous identified EMSs, including 55 WRs, 
at least two LBVs, and many OB supergiants. We can thus use these clusters to 
study the dependence of the evolution of the EMSs on initial stellar 
mass and the impact that such stars have on their environment~\citep{lan97,lan01}. 
In addition, dozens of emission-line stars 
have been identified outside these 
clusters~\citep{cot99,hom03,mun06,mik06,mau07,mau09,mau10a}. While some
of these stars were probably ejected from the clusters, others might form separately
in small groups or even in isolation~\citep{mau10a}. We have recently
carried out the first large-scale, high-resolution, narrow-band, 
Paschen-$\alpha$ survey, using the \nicm\ camera on the Hubble Space
Telescope (\hst ) \citep{wan10}. In addition to the
confirmation of the bulk of the known EMSs, the survey has led to the new 
detection of $\sim$ 80 stars with significant excess emission in the 
1.87 $\mu m$ band, which contains the H${\small \rm I}$ Paschen-$\alpha$ and several
He${\small \rm II}$ lines~\citep{don11}. These emission line
stars should mostly be EMSs with strong stellar winds; MS stars
and evolved low mass stars typically show absorption lines in their spectra. 
Indeed, follow-up spectroscopic observations have already confirmed 20 of
the stars as EMSs such as WRs, LBVs and OB supergiants~\citep{mau10b,mau10c};
most of the other emission-line stars are fainter and are yet to be
observed spectroscopically. By now, nearly 100 WR stars have been spectroscopically
identified in the central 200 pc~\citep{mau10c}. This sample is dominated by the late type
WN/WC stars. In contrast, Westerlund 1~\citep[$\sim$ 4.5 Myr,][]{cro06}, another
massive star formation region in the Galactic Disk, contains only 24
WR stars and half of the 16 WN stars are early-types (WN5-6). This
difference is consistent with the higher metallicity of the GC than
the Wd 1. Therefore,
the detections and identifications of the massive stars in the GC now
provide us with an excellent sample
of EMSs or their candidates to study their individual and statistical properties and to infer
relationships among the various types in a high metallicity environment. 
Furthermore, EMSs are prominent signposts for recent massive star formation.
From the study of the EMS populations in the GC, one can learn about the 
star formation mode, dynamics, and history in an extreme Galactic nuclear
environment, which is characterized by high gas 
temperature/density, strong magnetic field and tidal 
force~\citep{mor96}. 

Here we report a multiwavelength analysis of EMSs in the GC. 
We describe the multiwavelength data and the sample selection in \S~\ref{s:observation}, including the
new extinction correction and Paschen-$\alpha$ emission equivalent width (EW)
calculation of the sample stars. In \S~\ref{s:results}, we
present the near-IR, mid-IR and X-ray as well as emission line properties of the
stars. In \S~\ref{s:discussion}, we discuss the nature of the stars and their implications for the 
population, formation mode and history of massive stars. We summarize
our results in \S~\ref{s:summary}. Table 1 lists all the
acronyms and abbreviations used in this paper. 

\section{Sample Selection and Multiwavelength Data}\label{s:observation}
\subsection{Sample Selection}~\label{ss:sample}
Table~\ref{t:catalog} lists 180 sources included in the present study;
152 of them are from a primary list of the so-called Paschen-$\alpha$-emitting sources
(PESs) uniformly identified in the \hst/\nicm\ survey~\citep{don11}. 
The identifications are 
based on the large $1.87~\mu$m to $1.90~\mu$m flux ratios of these sources
($r=\frac{f_{F187N}}{f_{F190N}}$, $f_{F187N}$ and $f_{F190N}$ are the
intensities at 1.87 and 1.90 $\mu$m
, F187N, Paschen-$\alpha$, on the Paschen-$\alpha$ line, F190N, on the
adjacent continuum). This ratio describes the Paschen-$\alpha$ excess for each
individual star. For evolved, low-mass stars, which dominate 
the GC in a near-IR survey, 
this ratio is insensitive to the exact stellar type
 (with $<1$\% variation) and should instead trace foreground
extinction~\citep{don11}.  The local ratios ($\bar{r}$) towards 
different lines-of-sight are defined as the median average flux ratios
 of the closest 101 stars. All of the identified PESs (except for some
of the foreground stars, see section~\ref{s:fore}) have `significant
 Paschen-$\alpha$ excess', 
$N_s \equiv \frac{r-\overline{r}}{\sigma_{tot}}\geq4.5$,
where $\sigma_{tot}$ includes both
statistical and systematic photometric uncertainties of the ratio
measurements. $N_s$ is the signal-to-noise of the flux ratio of 
each source above the local ratio. Unfortunately, within the three clusters, this survey can
be problematic in isolating individual stars, because of the limited 
spatial resolution of the \nict\ camera used for the survey.  Its pixel size is 0.2\arcsec, which
undersamples the PSF (FWHM = ~0.17\arcsec\ at 1.90
$\mu$m). The large local background in the cluster cores further increases
the photometric uncertainty ($\sigma_{tot}$) and hence decreases 
the detection sensitivity, which may partly explain why more than
ten previously known WR stars in the Central cluster 
were not detected in our survey~\citep{don11}. Therefore, in~\citet{don11}, we also gave 
a secondary PES catalog, which includes the sources with
3.5$<$$N_s$$<$4.5. Compared to the primary catalog, this secondary 
catalog should include more spurious identifications. This catalog
 is only useful in the star clusters, in which we already know that 
there are many evolved massive stars, since some of the massive
 stars would not be included in the primary
 catalog because of the unusually high $\sigma_{tot}$ mentioned above~\citep{don11}.

Instead of using the secondary catalog to select
 the PESs in the three massive star clusters, we further 
refine the PES detection in the core regions of the Arches 
and Central clusters, utilizing 
archival \hst /\nics\ F187N and F190N observations (Programs \# 
7250 and 7364). By using the \nics\ data
with its smaller pixel scale (0.074\arcsec), we can improve
the photometric accuracy of the massive stars in these crowded regions. 
The basic parameters of these two sets of \nics\ observations 
are listed in Table~\ref{t:pre_obs}. We reduce the data
in the same way as was done in our survey~\citep{don11}. In
particular, we use the IDL program `Starfinder'~\citep{dio00} to
detect the sources and to obtain their photometry with the PSF
empirically extracted from bright stars within the same
observation. We use Eqn. (1) of~\citet{don11} to estimate the
photometric uncertainty of the detected sources, including the Poisson
fluctuations and the local background noise, as well as the uncertainty
of the photometric calibration of individual filters. The undersampling
uncertainty for \nics\ data is neglected, because it is expected to be
far less severe than for \nict\ data. With the
improvements in the detection limit, in addition to updating the 
photometries of the detected PESs (including
the separation of one source, ID 122, in Table 3 of ~\citealt{don11} 
into two separate stars, E48 and E51, as listed
in~\citealt{pau06}), we 
find three and ten new PESs in the Arches and
Central clusters, respectively (Table~\ref{t:catalog}), which are not
reported by~\citet[][their Table 3]{don11}.

We also include 14 WC stars within our survey region that were identified by other
spectroscopic observations, 
but are still not detected in our survey as significant PESs 
(MP1-MP14 in Table~\ref{t:catalog}; $N_s~< ~4.5$; references for these stars are listed there). 

The completeness of our survey has been discussed in our previous
paper~\citep{don11}. Briefly, the typical 50\% detection limit is
shown to be about 17 mag, although it can be as high as 15.5 mag 
in the Central cluster region. With the \nics\ data used in the
present work, the detection limit is further improved to 18.5 mag in 
the cluster regions. Because most of the 180 sources are significantly 
brighter than their local detection limits, we conclude that our 
EMS sample is nearly complete over the bulk of the survey field,
except in regions with exceptionally high extinctions. 

\subsection{Broad-band Near-Infrared Data}~\label{s:count_ir}

Three broad-band near-IR data sets are used: 1) HST
snapshot observations of the Arches, Quintuplet and Central clusters (Programs
7250, 7364, 7222 and 9457), 2)
\siri\ \citep{nag03}, and 3) \2mass\ \citep[][]{skr06}. 
The \hst\ observations were targeted on the
three clusters and their local background fields. 
Program 7364 utilized \nics\ to map the Arches and 
Quintuplet  clusters with three broad-band filters 
(F110W, F160W and F205W): the covered fields are 
1.4 ${\rm arcmin}^2$  (a total of 14 pointings) and 
2.0 ${\rm arcmin}^2$  (20 pointings), respectively. 
The Central cluster was mapped with one \nics\ (Program 7222) 
and four \nict\ pointings (Program7250) using F160W and F222M 
(covering a field of ~2.8 ${\rm arcmin}^2$). 
The medium filter (F222M) was used to avoid the saturation problem. 
There were also several \hst /\nics\ observations with F110W pointed toward a few
diverse fields within the Central cluster from Programs 7222 and 9457.
All of our PESs in the three clusters have counterparts in these
broad-band observations. But they cover only about 1.5\% of our \hst /\nicm\
Paschen-$\alpha$ GC survey field. 

We thus use the \siri\ and \2mass\ catalogues for those PESs 
not covered by the \hst\ broad-band observations \footnote{Simultaneous 3-color InfraRed Imager for Unbiased Surveys (SIRIUS) was taken by the Infrared Survey Facility (IRSF) 
in South Africa, with a pixel scale of 0.45\arcsec~\citep{nag03}. 
The survey includes the region $|l|$ $<$ 2 degree and $|b|$ $<$ 1
degree, with an angular resolution $\sim$ 1.2\arcsec\ in the J band, 
better than that of 2MASS, $\sim$2\arcsec. }.
\siri\ covers our entire sample field in J (centered at 1.25 $\mu$m),
H (1.63 $\mu$m) 
and K$_s$ (2.14 $\mu$m). For a few bright PESs, which are
saturated in \siri, we adopt the corresponding broad-band
magnitudes (J: centered at 1.24 $\mu$m, H: 1.66 $\mu$m and K$_s$: 2.16 $\mu$m) provided by
\2mass\ \citep{skr06}. We identify the closest broad-band source as
the counterpart of each PES
with a matching radius of 0.1\arcsec. We also examine the individual
matches by eye to make sure that the counterparts between the three catalogs are real matches. 

We convert the photometries of these catalogues into a common standard to facilitate a systematic analysis.
Fig.~\ref{f:trans_curve} shows the transmission curves of the
different systems in various filters. The filter response curves of
the \hst\ are much wider than those of the ground-based
systems. We obtain approximate magnitude conversions using
counterparts, or the closest sources in two different catalogs with offsets
 less than 0.2\arcsec. For \hst\ and \siri , 
considering their much different filter systems, we first 
use a very crude match selection criterion: 
each source pair has a K-band magnitude difference less 
than 2 magnitudes. The H and K$_s$ filters of \siri\ and \2mass\ 
are very similar; we thus use a tighter criterion, requiring the 
K$_s$ magnitude difference to be less than 0.3.
We then utilize least-squares fits to obtain the 
first-order color-dependent conversions to the \siri\ magnitude 
system, and a $3-\sigma$ clipping to remove the outliers. 
This $3-\sigma$ clipping effectively removes the \siri\ sources 
with multi \hst\ counterparts within the search radius, 
even in crowded regions such as the Central cluster. Finally, we use 
least-squares fits again on the remaining source pairs to derive the following conversions:
\begin{eqnarray}
J_{siri}=J_{HST}-0.079(\pm0.006)(J_{HST}-K_{HST,F205W})-0.181(\pm0.030)\\
H_{siri}=H_{HST}-0.011(\pm0.015)(H_{HST}-K_{HST,F205W})-0.228(\pm0.024)\\
K_{siri}=K_{HST,F205W}-0.021(\pm0.016)(H_{HST}-K_{HST,F205W})-0.435(\pm0.026)\\
J_{siri}=J_{HST}-0.159(\pm0.004)(J_{HST}-K_{HST,F222M})-0.247(\pm0.031)\\
H_{siri}=H_{HST}-0.160(\pm0.004)(H_{HST}-K_{HST,F222M})-0.246(\pm0.011)\\
K_{siri}=K_{HST,F222M}+0.068(\pm0.003)(H_{HST}-K_{HST,F222M})-0.562(\pm0.007)\\
J_{siri}=J_{2mass}+0.043(\pm0.001)(J_{2mass}-K_{2mass})-0.021(\pm0.003)\\
H_{siri}=H_{2mass}-0.001(\pm0.001)(H_{2mass}-K_{2mass})+0.075(\pm0.002)\\
K_{siri}=K_{2mass}+0.070(\pm0.001)(H_{2mass}-K_{2mass})-0.075(\pm0.002)
\end{eqnarray}
Fig.~\ref{f:mag_transfer} compares the \hst /\nicm , \2mass\ and \siri\ magnitudes 
before and after these conversions. The numbers of sources used to
derive the equations above are also marked in each panel of Fig.~\ref{f:mag_transfer}. 
Table~\ref{t:catalog} includes the J, H and K$_s$ magnitudes of the
PESs, all in the \siri\ photometric system. 

\subsection{Extinction Corrections}~\label{s:extinction}
Using the broad-band near-IR magnitude measurements, we improve our
extinction correction for the PESs. In contrast to the use of 
the statistically constructed extinction map detailed in~\citet{don11}, 
here we directly estimate the extinction
along the sightlines toward individual PESs, based
on the J-H and/or H-K$_s$ colors, if available. 

Table~\ref{t:color} lists the intrinsic colors, (J-H)$_{0}$ and 
(H-K$_s$)$_{0}$, adopted for the PESs. 
The large dispersion in
the (H-K$_s$)$_{0}$ distribution of WR stars is mainly because their K$_s$-band 
intensities can be substantially contaminated by the free-free
emission from their strong stellar winds and/or the surrounding 
dust thermal emission. For those PESs with no available spectroscopic 
identifications, we simply adopt the mean values obtained for O
[(J-H)$_{0}=$-0.11 and  (H-K$_s$)$_{0}=$-0.1] and WR stars
[(J-H)$_{0}$=0.02$\pm0.056$ and  (H-K$_s$)$_{0}=$0.19$\pm0.182$] (see Table.~\ref{t:color}): 
(J-H)$_{0}=$-0.045$\pm0.121$ and (H-K$_s$)$_{0}=$ 0.045$\pm0.327$, the 
uncertainties of which include the color differences of O and WR stars.

Following \citet{nis06}, we adopt the relative extinction coefficients for the broad-band filters (J, H and K$_s$) of \siri\  toward the GC as
$A_J$:$A_H$:$A_{Ks}$=1:0.573:0.331. Assuming a single power
law for the extinction curve between H and K$_s$, we then obtain the
slope, $\alpha$=2.02, $A_{F190N}$=1.271$A_{Ks}$,
$A_{F187N}$=1.306$A_{Ks}$ and 
\begin{eqnarray}
A_{F190N}=0.985(A_J-A_H)\label{e:A_J-H}\\
A_{F190N}=1.738(A_H-A_{Ks})\label{e:A_H-K} \\
A_{F190N}=90.56\times {\rm log}
(\frac{1.015}{\overline{r}})\label{e:A_r}.
\end{eqnarray}
The definition of $\overline{r}$ was given in \S~\ref{ss:sample}. 

Fig.~\ref{f:av_JHK_compare}a compares the $A_{F190N}$ values obtained from
Eqns.~\ref{e:A_J-H} and~\ref{e:A_H-K}
for the PESs with available J, H and K$_s$ measurements and
known stellar types. The mean difference between $A_{F190N}$(J-H)
and $A_{F190N}$(H-K$_s$) is 0.01$\pm$0.03 for OB supergiants,
0.02$\pm$0.04 for WN stars, and 0.16$\pm$0.09 for WC stars. 
For WC stars, $A_{F190N}$(H-K$_s$) is overestimated because of the
contamination by hot dust emission (\S~\ref{s:mid}) in K$_s$. Therefore, 
we use only $A_{F190N}$(J-H) to represent the foreground extinction of
WC stars. For 
other PESs, we generally adopt the mean of  $A_{F190N}$(J-H) and 
$A_{F190N}$(H-K$_s$). But for PESs with no J measurement, we 
adopt $A_{F190N}$(H-K$_s$), if available; otherwise, we just use the extinction 
from Eq.~\ref{e:A_r}, which represents the local average extinction~\citep{don11}. In Fig.~\ref{f:av_JHK_compare}b, we compare the
$A_{F190N}$ values derived from Eq.~\ref{e:A_J-H} and Eq.~\ref{e:A_r}
for the PESs with available J and H magnitudes, excluding the ten
potential foreground stars identified in \S~\ref{s:fore}. The similar extinctions
derived from these two equations
($A_{F190N}(\bar{r})$-$A_{F190N}$(J-H)=-0.1$\pm$0.07)
 suggest that most of these PESs are
indeed within the GC. The 
extinctions so obtained are listed in Table~\ref{t:catalog}. 

\subsection{Paschen-$\alpha$ Equivalent Widths}\label{s:data_ew} 
We recalculate the Paschen-$\alpha$ equivalent widths (EWs) 
for those stars with improved 
magnitude and extinction measurements. The calculations follow
the formula:\\
\begin{equation}\label{e:ew}
EW_{1.87\mu
m}=\frac{f^o_{F187N}-1.015*f^o_{F190N}}{1.015*f^o_{F190N}}*\delta\lambda,
\end{equation}
where $f^o_{F187N}$ and $f^o_{F190N}$ are the extinction-corrected
fluxes of the PESs 
in the F187N and F190N bands, $\delta\lambda\sim$0.191 $\AA$ is the 
effective FWHM
of the F187N filter, and 1.015 
represents a typical F187N to F190N continuum flux ratio, which is obtained
for a K0III star and is insensitive to the exact stellar type, 
as long as there is no absorption lines in F187N~\citep{don11}. The EWs for the PESs are
 listed in Table~\ref{t:catalog}. 

\subsection{Mid-Infrared Data}~\label{s:midir}
To further examine the dust emission from the PESs, particularly WC stars, we 
include the mid-IR measurements from the Spitzer/IRAC GALCEN survey
~\citep[][]{sto06}. We use a search radius of 1\arcsec\ to find the IRAC counterpart
of each PES. Because of the limited spatial resolution of the
IRAC camera, most of the EMSs within the cores of the Arches and
Central clusters do not have mid-IR measurements. 
The mid-IR magnitudes of the PESs in the 3.6, 4.5, 5.4 
and 8.0 $\mu$m bands are included in
Table~\ref{t:catalog}. The photometric uncertainties in the GALCEN source 
catalog are systematically underestimated by a factor of about
 5 in the 3.6 and 4.5 \micron\ bands\footnote{see
  http://www.astro.wisc.edu/sirtf/glm2\_galcen\_comparison.pdf}. Therefore, we correct for the uncertainties from the GALCEN catalog by this factor.

\subsection{Spectral Energy Distribution Analysis}~\label{s:SED}
We conduct a simple spectral energy distribution (SED) fit to
characterize the dust emission typically associated with WC stars. 
In our sample, 16 WC stars have available photometries in
the six bands: $m_{J}$, $m_{H}$, $m_{K_s}$, $m_{F190N}$, [3.6] and [4.5]. 
We adopt the extinction laws provided by~\citet{nis06} and~\citet{ind05},
to correct for the extinction in each of the 
six bands,
scaled to $A_{F190N}$ 
($A_J$: $A_H$: $A_{Ks}$: $A_{3.6~\mu m}$: $A_{4.5~\mu
  m}$=2.38:1.36:0.79:0.43:0.34). 
Zero points of 835.6, 280.9, 179.9 Jy are used to convert the magnitude into flux for 
F190N, [3.6] and
[4.5].\footnote{http://www.stsci.edu/hst/nicmos/performance/photometry
  and \spit /\irac\ instrument handbook, Table 4.1} As suggested
by~\citet{mau10a}, because the photometry of \siri\ and \2mass\
are similar~\citep{nis08}, the zero points for the J, H and K$_s$
magnitudes of the \2mass\ system are used for sources in the \siri\
catalog (for which we find no zero point information). 
Five of these WC stars (P28, P34, P60, P94,
P101) have relatively blue colors (K$_s$-[3.6]$<$2)
and do not show evidence for 
surrounding dust (those without 'd' in their type definition;
Table~\ref{t:catalog}). We first normalize the flux measurements of
each of these WC stars to $f_{J}=1~Jy$ and then
median-average them to construct a ``dust-free'' SED template, which should 
include any potential free-free emission from the stellar winds as well as
the emission emerging from the stellar atmospheres of these
sources. The uncertainties of the median-average SED in the six bands 
are less than 20\%. We fit the SED of the other 11 WC stars with this template, plus a diluted
blackbody to characterize hot dust emission with an emissivity
proportional to $\lambda^{-1}$
, $I_{\nu}$ =$\frac{a}{\lambda^4}\frac{1}{exp(\frac{b}{\lambda})-1}$,
in units of Jy, where $a$ is the normalization factor and
$b=\frac{hc}{kT}$~\citep{gou95} (see Fig.~\ref{f:SED_fitting}). 

\subsection{X-ray Measurements}~\label{s:count_x}
We utilize the measurements made by the deep \chan\ ACIS
survey~\citep{mun09} to search for X-ray counterparts of the
PESs. The matching radius used
is the smaller of 3$\sigma_{x}$ (where $\sigma_{x}$ is
the X-ray source positional uncertainty) and
2$\arcsec$ (which is used to reduce the source 
confusion). To estimate the probability for chance positional
coincidences, we repeat the same matching procedure after each of
eight systematic shifts of the X-ray source positions (i.e., the
matches were actually done on a grid of
$\delta RA$ $\subset$ [-5\arcsec, 0\arcsec, 5\arcsec],
$\delta Dec$ $\subset$ [-5\arcsec, 0\arcsec, 5\arcsec]). The mean number of
chance positional coincidences in the eight shifts is found to be 7.5,
which mostly occur in the core regions of the three clusters. Outside
the clusters, the number of chance positional coincidences is only
1.8. 

Table~\ref{t:pal_x} lists 35 PESs with X-ray
source counterparts. In addition to the source IDs listed
in~\citet{mun09}, the table also includes X-ray source positional
uncertainty ($\sigma_x$), 0.5-8 keV fluxes
and hardness ratios (see the note to the table). P122a and P122b, 
members of the IRS 13 compact stellar complex~\citep{mai04}, are just
0.3\arcsec\ apart (compared to the \chan\ on-axis FWHM
$\sim$0.5\arcsec) and are listed to have the same
X-ray source counterpart.  

Most of these PES/X-ray source matches should represent genuine
counterparts. Table~\ref{t:pal_x} includes the percent probability (P)
for each PES/X-ray source match to be genuine. The probability was
calculated  by Mauerhan et al. (2009) using a match radius 
that is the quadrature sum of the positional uncertainties of the
\chan\ and \siri\ measurements. Those matches studied
previously~\citep{mun06,mik06,mau10a} all have high probabilities ($>$
20 \%), except for P147 (partly because of its large positional
uncertainty: 0.9\arcsec). Because of saturation in the \siri\
observation, the source P38, which was identified in our \hst\
survey~\citep{don11}, was not listed in Table 3 of~\citet{mau09}, and
therefore does not have an existing `P' value.~\citet{mau09} also 
did not match any X-ray sources to the four
PESs in the Central cluster (P123, P125, P158 and P161 in
Table~\ref{t:pal_x}), because the \siri\ catalog suffers from
confusion in this crowded region. However, these sources are known
WR stars~\citep{pau06}. Four other PES/X-ray sources (P105, P108, P133 and P139) have low `P'
($<$15\%) and have not yet been spectroscopically
classified. Therefore, whether they are real matches remains uncertain
(see also \S~\ref{s:discussion}).

\section{Results}\label{s:results}

The locations of our 180 sample PESs are marked on the 
Paschen-$\alpha$ mosaic image of the GC in Fig.~\ref{f:pal_sou}.
A total of 30, 17 and 33 of the sources are located within 2$r_{c}$ of 
the Quintuplet, Arches and Central clusters, respectively;
the cluster radius, $r_c$, of each cluster is from Table 5
of \citet{fig99a} (Quintuplet: 1pc, Arches: 0.19 pc, Central: 0.23
pc)  \footnote{The core radii of the Arches and Central clusters are 
consistent with recent work: Arches: 0.14$\pm$0.05 pc~\citep{esp09}
and Central: 0.22$\pm$0.04 pc~\citep{sch07}. The tidal radius ($\sim$1
pc) of the Arches cluster derived from 
dynamical simulations~\citep{kim99,por02} is larger. Using 
this latter radius, two additional PESs (P17 and P79) 
would be classified to be within the Arches cluster, 
which, however, would not qualitatively affect our results and conclusions.}. We refer to these sources as `cluster' PESs and to
the remaining
100 PESs as `field' ones.

\subsection{Foreground Extinction Properties}
The extinction distributions of the PESs in
different regions are depicted in Fig.~\ref{f:av_dis}. The means (standard
deviations) of $A_{F190N}$ are 3.32(0.56), 3.20(0.24), 3.71(0.80) and 2.98(0.89)
for the Quintuplet, Arches, Central and `field' PESs, respectively. The relatively large
dispersion for the Central cluster PESs is largely due to the 
exceptionally large extinctions of MP7 (E31, WC9) and MP10 (E58,
WC5/6, with no
available J-band measurement; Table~\ref{t:catalog}), most likely due to the extra 
extinction of their circumstellar dust. 
Excluding these two sources reduces the mean and standard deviation 
to 3.55 and 0.5. In general, the PESs within the individual clusters have very
similar extinctions. The corresponding $A_{Ks}$ (standard deviation) for the 
Quintuplet, Arches and Central clusters are 2.61(0.44), 2.52(0.19) and
2.79(0.39), which are consistent with existing measurements:
3.1~\citep[0.5,][]{lie10}, 2.13-4.14~\citep{esp09} and
2.54~\citep[0.12,][]{sch10}, respectively. 

Fig.~\ref{f:IR_color} shows the J-H vs. H-K$_s$ and K$_s$ vs. H-K$_s$
diagrams. 
The color-color plot contains 119 PESs
with both J-H and H-K$_s$ measurements, while the color-magnitude diagram is used for
the 45 PESs with only H-K$_s$ colors. Most of the PESs are located in the ranges
of H-K$_s$$\in[1,3]$ and J-H$\in[2,4]$. In Fig.~\ref{f:IR_color}, we also plot all of the stars 
in the \siri\ catalog, which are mainly low mass giants, plus
foreground MS stars. They should typically have similar intrinsic near-IR
colors (on the Rayleigh-Jeans side of the SED). Therefore, 
the observed colors of such stars should
trace the expected reddening path. In contrast, the PESs show 
colors that can be significantly redder than the path 
(e.g., up to 1-2 mag in H-K$_s$), apparently a result of 
enhanced emission at long wavelengths, due to
 free-free and/or hot dust emission. The remaining 
16 PESs do not have H or K$_s$ measurements. 

\subsection{Identification of Foreground Stars}\label{s:fore}
The above extinction properties of the PESs can help us 
identify/confirm stars that are in the foreground of the GC. 
We find that nine stars have H-K$_s$ $<$ 1 and J-H $<2$ 
(if available), including the two foreground OB
supergiants suggested by~\citet{mau10c}.
Table~\ref{t:fore} includes the extinction-corrected F190N magnitudes 
and updated $N_{s}$ values (see \S~\ref{ss:sample}) of these foreground PESs. 
Four of them (P27, P102, P140 and P149) 
still have significant F187N excesses, with $N_s~>~4.5$. 
Although the $N_s$ values of the remaining five PESs are now $<4.5$,
they may still be foreground
emission-line stars. For example,
P38, with the second smallest revised $N_{s}$,  
spectroscopically exhibits several emission lines around 2.112-2.115 
$\mu$m and is identified by~\citet{mau10c} as an O4-6I star 
at a distance of 3.6 kpc. 

The nine foreground PESs can be divided into two
groups according to their extinction-corrected F190N magnitudes. The first
consists of P38, P140 and P145, which  have comparable foreground
extinctions and intrinsic brightnesses. As both P38 and P140 are EMSs,
we suggest that P145 is likely to be as well.
The other six PESs form the second group. They are a factor of at
least $\sim 10$ fainter and have comparable 
or smaller (but within a factor of $\sim 3$) extinctions, compared to
those in the first class, indicating 
that they are typically closer and substantially dimmer by more than
$>~2.5$ mag than P38, for example. According to~\citet{fig95}
and~\citet{cro06}, even the dimmest WR ($M_{Ks}\sim$-2.63) is no 
more than a factor of 10 fainter than the first group sources.
 Therefore, we suspect that the PESs in this second 
class are not EMSs (such as WR and OB supergiants) and are 
probably cataclysmic variables (CVs), which often show hydrogen
emission lines~\citep{dhi95}. 
CVs usually contain low-mass MS secondaries ($\lesssim 1 M_\odot$)
with M$_{Ks}$ in the 2-9 mag range~\citep[Fig. 9 of ][]{hoa02}. 
 Therefore, with the extinction taken into account, only relatively
 nearby CVs (up to a distance of $\sim 4$ kpc) can be detected in our survey.

When X-ray counterparts are available, their spectral properties can also be used to check the consistency of the 
identifications and even to provide clues to the nature of the sources. 
Table~\ref{t:pal_x} summarizes the 35 matches between the PESs and known X-ray sources.
Most of these matches have been examined with previous ground-based 
near-IR observations, and have been suggested to be massive colliding-wind binaries 
within the GC (except for P38; \citealt{mau10c}). 
The X-ray hardness ratio HR0 is often used to
distinguish foreground sources from those at the
GC (e.g.,~\citealt{mau10a,mau10c}).~\citet{mun09} suggested that a source
 with HR0 $\lesssim$ -0.175 probably has a distance less
than 4 kpc. There are four such EMSs, including P38, consistent
with its foreground star identification. 
P145 also has a soft spectrum, consistent with being a 
foreground massive star, according to its near-IR color above. 
The statistical uncertainties in the HR0
values for P109 and P138 are too large to provide useful
constraints. For example, the red near-IR color of 
P109 ($H-K_s=1.6$) indicates that it could
be within the GC, instead of a foreground source. 

The nine stars with blue near-IR colors in Table~\ref{t:fore}, plus
one soft X-ray star (P138), are excluded
from the discussion to follow, which focuses on the EMSs within the GC.

\subsection{Continuum and Paschen-$\alpha$ Emission Properties of GC
  EMSs}\label{s:lum}
We present the F190N magnitude distributions of our 170 GC EMSs in
Fig.~\ref{f:lum_pal}. The distribution becomes increasingly broad 
from the Arches ($\sim$2.5 Myr old), Quintuplet ($\sim$4 Myr old), to 
Central ($\sim$6 Myr old) clusters. There are also systematic magnitude 
differences among the distributions,  which may be characterized by their 
mean F190N magnitude (standard deviation): $\sim$-6.21(0.58),
$\sim$-6.93(2.09) and $\sim$-5.26(1.33) for the Arches, Quintuplet 
and Central clusters, respectively. The unusually large displacement of 
the Quintuplet cluster from the other distributions may be due to the presence of several
very bright stars ($M_{F190N} \lesssim -8$) 
due to substantial contamination by circumstellar dust 
emission: mainly one LBV and the five Quintuplet
proper members (QPMs, dusty WCs), which have unusually 
red colors~\citep{fig99a,lie09}. 
Excluding these six stars, the mean would decrease to
$\sim$-6.18(1.55). The standard deviation further decreases to 0.85,
after the three dim stars with $M_{F190N} > -5$ are removed. On the 
other hand, the  stars in the Central cluster
are fainter by $\sim$ 1-2 magnitudes. This offset is consistent with
the results derived from the $K_{s}$
observations~\citep[][]{fig02,pau06} 
and is at least partly due to the larger age of the Central
cluster. Therefore, the magnitude distribution shift 
in general seems to be linked to the age of a cluster. 

The magnitude distribution of `field' sources spans a wide range and is possibly
double-peaked. About 50\% (46) of the `field' sources have absolute
1.90 $\mu$m magnitudes similar to those in the three young clusters,
while the other half are considerably fainter. Compared to those in
the bright peak, the mean extinction of
the PESs in the dimmer peak is smaller by $\sim$ 
0.4 magnitude, which cannot explain the large magnitude difference
between these two peaks ($>$ 2 mag).

Fig.~\ref{f:ew} presents an EW vs. $f^o_{F190N}$ plot
for the PESs with identified stellar types in different regions. The stellar types 
are identified from previous ground-based spectroscopic
observations (Table~\ref{t:catalog}). WNL stars in the Arches cluster
(triangles) show an apparent linear relation between the EW and 
the stellar luminosity, as already demonstrated
by~\citet{fig02}. They pointed out that this relation is expected 
from radiation-driven stellar winds. A similar linear relationship 
seems to hold for WNL stars
in the Central cluster, excluding the Ofpe/WN9 stars (filled squares). But 
these WNL stars are systematically dimmer by a factor of $\sim$ 5 than those in the Arches
cluster (see \S~\ref{s:lum}
and Fig.~\ref{f:lum_pal}), implying that the EW  is
determined by not only the luminosity, but also the stellar age.
The three WNL stars in the Quintuplet cluster are more similar to those in
the Arches cluster, than to the Central cluster. In Fig.~\ref{f:ew}, 
most of `field' WNL stars (`star' symbol) are
located between the Arches and Central clusters. However, `Ofpe/WN9' stars seem to be distributed
differently and have substantially smaller EWs ($<100\AA$) than 
WNL stars of similar
$f^o_{F190N}$. Both the Quintuplet and Central clusters are rich in WC stars, with a broad
EW range. The three WC stars in the Quintuplet with EW $>$ 250 $\AA$ (P12,
P13, P60, which are all WC8) have strong He{\small \rm I}
(2.059 $\mu$m) and He{\small \rm II} (2.19 $\mu$m) in their near-IR K$_s$-band
spectra~\citep{fig99a,lie09}. 
 But many WC stars have very weak line emission,
which explains why 14 WC stars are not identified in our PES survey
(see \S~\ref{s:observation}). Five of them are the QPMs,
indicating that their stellar photospheres are completely blanketed by dust.
 Most of the OB supergiants also have very small EWs, indicative of their
 weaker stellar winds. 

In the three clusters, no EMS is detected to have 
$f^o_{F190N}<2\times10^4~\mu Jy$ and EW$<100\AA$ at the same time. This is likely
due to the relatively high flux detection limit in the cluster fields due to crowding. 
Fig.~\ref{f:f_uf} shows
the photometric uncertainty $\sigma_{tot}$ (see \S~\ref{ss:sample}) vs. $f^o_{F190N}$ distribution
of the detected sources within the three clusters, compared with the unclassified
`field' EMSs, which have low $f^o_{F190N}$ and EWs (we use
$f^o_{F190N}$ instead of $f_{F190N}$ to avoid the effects of the differential
foreground extinction). 
Because of the high local stellar background, 
the photometric uncertainty in the Central cluster is much higher than those in
the Arches and Quintuplet clusters. 
The photometric uncertainty for the Quintuplet cluster (which only 
has the low resolution of NICMOS/NIC3 F187N/F190N data)
is also larger than that in the Arches cluster (for which NIC2 data are 
used; \S~2.1). Therefore, the non-detection of the analogue of the
field EMSs, which have relatively low brightness and EWs, in the Quintuplet and Central clusters could
be due to their large $\sigma_{tot}$. In the case of the young Arches
cluster, however, massive stars are still mostly on the MS, except for 
a few very massive ones
($\geq 60M_{\odot}$) that may have just turned into bright WNL stars or OB supergiants. 

\subsection{Dust Emission in the Mid-IR}\label{s:mid}
We present a color-color diagram for the 36 EMSs with available
H, K$_s$ and [8.0] magnitude measurements in Fig.~\ref{f:Mid_color}. Limited by the
angular resolution of Spitzer/IRAC ($\sim$2\arcsec), these PESs are
all in the `field', although a few of them are located at the peripheries 
of the Quintuplet and
Arches clusters. Most of the WN stars and OB supergiants are only slightly redder in
K$_s$-[8.0] than average field stars.
The three OB supergiants (P35, P100, P112) that have 
unusually red color ($K_s-[8.0]>5$) are all associated with
local, strong, extended Paschen-$\alpha$ features, so their [8.0] flux
density is presumably dominated by Polycyclic 
Aromatic Hydrocarbon (PAH) emission. While all WN stars are
constrained to [2,4], a significant number of WC stars have very red
K$_s$-[8.0] color. Six
of the nine WC stars with red K$_s$-[8] ($>$3)
in Fig.~\ref{f:Mid_color} have been suggested to be associated with
hot dust~\citep{mau10a,mau10c,lie09,fig99a}. 

Fig.~\ref{f:Mid_color_1} shows an EW vs. K$_s$-[3.6] plot for 68 EMSs. We can see that WNL stars/OB supergiants and WC stars seem to follow different
trends. For WNL stars, the EW increases with K$_s$-[3.6]. But, the 
trend is opposite for WC stars, at least for those with low EWs. The EWs of
the OB supergiants show a very weak, possibly positive dependence on
K$_s$-[3.6]. 
These different trends probably indicate different relative 
contributions from various sources of emission. The free-free emission from the 
strong winds of the WNL stars/OB supergiants should
dominate in the mid-infrared~\citep{wri75}, resulting in the positive
EW vs. K$_s$-[3.6] correlation. In contrast, the strong dust emission 
from WC stars may dilute the Paschen-$\alpha$ line
emission and dominate the mid-infrared. 
For example, QPMs 
have very red infrared spectra and appear nearly
featureless. Indeed, dusty spirals have been resolved spatially around two 
of these WC stars based on high-resolution adaptive optics
observations~\citep{tut06}. The enhancement of the hot dust emission
at the longer wavelengths suppresses the line contrasts. As a result, a significant fraction
of WC stars may thus not be detected as PESs.

The results from the SED fits to the 11 WC stars (\S~\ref{s:SED}) 
with enhanced mid-IR emission are presented in
Table~\ref{t:dust} and Fig.~\ref{f:SED_fitting}. The combination of the
``dust-free'' template and the diluted dust blackbody fits the bulk of the
WC stars well. For MP1, we notice that the Spitzer/IRAC
source catalog includes an additional source that is only 1\arcsec\ 
away and is detected only at 4.5 $\mu$m, but not at 3.6 $\mu$m. 
This source may thus represent a spurious detection and may have caused 
an underestimate of the intensity of MP1 at 4.5
$\mu$m in the Spitzer/IRAC catalog. Therefore, we 
have excluded the 4.5 $\mu$m data point from the SED fitting of MP1 in
Table~\ref{t:dust} and Fig.~\ref{f:SED_fitting}. Both the
table and figure show that the dust
contribution at 1.90 $\mu$m varies significantly from
one source to another and reaches 
more than 30\% for the QPMs. After removing the dust contribution 
from the F187N and F190N flux densities, we recalculate the EWs, 
the results of which are also listed in Table~\ref{t:dust}. For the WC stars not detected in our
survey (MP*), their EWs are increased, though still very small
($<$30$\AA$). Since dusty WC stars 
are presumably formed within colliding-wind
binary systems (\citealt{cro07}, see also~\citealt{tut06}), the presence of massive (OB MS) companions
 with flat spectra and even Paschen-$\alpha$ absorption lines could
 also account for the low EWs. The fitted dust
temperatures are in the range 550-1000 K. 
We derive the dust mass in these WC stars (see Table.~\ref{t:dust}) using Eqn. 2
of~\citet{gou95}:
$M_d$=$5.1\times10^{-11}S_{\nu}D^2\lambda^4_{\mu}(e^{1.44\times10^4/
  \lambda_{\mu}T_d}-1)~M_{\odot}$, where $S_{\nu}$ (in units of mJy)
and D (Mpc) are the specific blackbody flux at frequency $\nu$ and
the distance to the source, respectively. In
order to exclude the contamination from the stellar atmosphere at 
short wavelengths, we use
$S_{\nu}$ at 4.5 $\micron$, except for MP1 (at 3.6
$\micron$). The dust mass in each of the five QPMs (MP1, MP2, MP3, MP5 and
MP6) is usually large, $10^{-6}-10^{-5}~M_{\odot}$, while the other six
WC stars have dust masses of order $10^{-8}-10^{-7}~M_{\odot}$, similar to the WC stars studied
in the solar neighbourhood~\citep{vee98,mar99}. 

\section{Discussion}\label{s:discussion}
In this paper, we construct a catalog of 180 EMSs in the Galactic center region,
 including many identified WN, WC, LBV and OB supergiants. Based on the
brightness and EW distributions of such stars, as discussed
in~\citet{mau10c}, this catalog should include nearly all of the
WN stars in our survey area, while a significant fraction of faint WC
stars may still be missing. Even in the core regions of the Arches and
Central clusters, while the Paschen-$\alpha$ survey itself may be
quite incomplete because of the crowding and nebulosity, the
dedicated \nics\ observations of higher spatial resolution and
ground-based spectroscopic observations should have identified the bulk
(if not all) of the WN stars. 

Based on the above results, we can gain some insight into the nature of 
those still unclassified PESs, especially those outside the clusters.
 Fig.~\ref{f:pal_all} compares the PESs with and without
 spectroscopic classifications in the  EW vs. $f^o_{1.90~\mu m}$ plot. 
For ease of the 
discussion, we empirically divide the plot into three parts,
 each of which is populated mainly  by one single stellar type:
the WNL-dominated region contains 67\% of the WNL stars, 
the OB supergiant-dominated region contains 96\% of the OB
supergiants, and the WC-dominated region contains 42\% of the WC stars. 53\%
of the WC stars with very small EWs fall into the `OB supergiant'
region. In Fig.~\ref{f:Mid_color_1}, we find that the WC stars with
red color (larger K$_s$-[3.6]) tend to have smaller EWs. Therefore, we propose that
the SEDs of WC stars falling into the `OB supergiant' region could be substantially 
contaminated by the thermal emission of hot circumstellar dust, which
may affect the flux even at 1.90 $\mu$m.

The unclassified PESs have
either low EWs and/or low intrinsic brightnesses. 54 out of
67 such PESs are within the `OB supergiant' region
(Fig.~\ref{f:pal_all}). One in the Central cluster with EW=$343\AA$ falls 
within a tight cluster of WNE stars in
Fig.~\ref{f:pal_all}, and is detected
in the \hst /\nics\ observations, but not in the
\hst /\nict\ Paschen-$\alpha$ Survey~\citep{don11}, because of the
confusion limit in the very crowded region. The other 12 unclassified PESs are all in the `WC' region. 

The unclassified PESs are scattered around in the plots of Fig.~\ref{f:Mid_color} and
Fig.~\ref{f:Mid_color_1}. Their MIR properties suggest that they
include sources from various subgroups, such as dusty WC stars and OB supergiants. 
For example, P44, P115 and P141 are very red ($K_s-[8.0]>4$ and
$K_s-[3.6]>3$) and are unrelated to any local diffuse emission in
the Paschen-$\alpha$ map. Therefore, they are most likely 
dusty WC stars. Several other unclassified PESs fall into the
 `OB supergiant' region. Spectroscopic identification is
 needed to determine their exact stellar types. 
 
The six previously unclassified PES/X-ray source matches (Table~\ref{t:pal_x})
are all located outside
the three clusters. 
They have 0.5-8 keV 
fluxes of $\sim$ 4$\times10^{-7}$ photons cm$^{-2}$
s$^{-1}$, dimmer than most of the other X-ray
sources in Table~\ref{t:pal_x}. 
Because of the low signal-to-noise ratios of the sources, 
we only use their X-ray colors for spectral characterization. In \S~\ref{s:fore}, HR0 values
are used to check the consistency of classifying P138 and P145 as foreground stars. 
Fig.~\ref{f:X_color} further presents the relationship between the HR2 and the
photon fluxes of these X-ray counterparts. Unlike HR0,
HR2 is a measure of the spectral characteristics at relatively high energies,
and is less affected by foreground
extinction. As stated in~\citet{mau10a}, the X-ray counterparts
of  the EMSs are mostly softer and brighter than the majority of the X-ray
point sources toward the GC, which are typically 
CVs and have relatively hard spectra, with thermal
temperature $\gtrsim$ 2 keV. Among the six
new detections, P105 and P133 have HR2 ($\lesssim$ 0.4) similar to the other
known EMSs with X-ray counterparts, suggesting that they may have a similar
origin. On the other hand, P108 and P139, with relatively 
hard HR2 ($>$ 0.5), fall into the CV region in Fig.~\ref{f:X_color}. 
However, as we mentioned in \S~\ref{s:fore}, after correcting for 
the distance modulus and the extinction~\citep[$A_K$=2.22,][]{don11}, 
the CVs in the GC should be dimmer than 18.7 magnitude, 
which is well below the source detection limit of our
survey~\citep[see Fig. 14. of ][]{don11}. 
Therefore, P108 and P139 are probably unrelated to
 the X-ray point sources, consistent with their poor 
matching probability (Table~\ref{t:pal_x}).

The number of `field' EMSs is comparable to that within the three
clusters~\citep{mau08,don11}. Fig.~\ref{f:lum_pal} shows that the absolute
magnitude distribution of these `field' EMSs is much wider than 
those of individual clusters and shows two peaks, perhaps suggesting
 two distinct episodes of star formation. The absolute magnitudes of more than half of the EMSs 
fall between the peaks of the Arches and Central clusters. Some of these 
EMSs are WNLs and are mostly located between the loci defined by 
WNLs within the Arches and Central clusters in the EW vs. F190N plot
of Fig.~\ref{f:ew}. These 
facts suggest that these EMSs have ages between those of the Arches
 and Central clusters. The remaining fainter unclassified `field' PESs
 are presumably older. Some of them have magnitudes comparable
 to those PESs in the Central cluster that are WC stars 
($M_{Ks}\in [-3,-5]$). Therefore, some of these unclassified `field'
 PESs could also be WCs and trace the stellar population with ages 
comparable to that of the Central cluster.

Although some fraction of the field EMSs may represent the EMSs that have been
ejected from the three clusters via dynamical processes, at least some,
especially those young ones, likely formed in small groups or even in isolation.
Probably the most convincing evidence for this scenario is the apparent
associations of a few of EMSs with local H{\small \rm II} regions, such as H1,
H2~\citep{zha93}, and several others in the Galactic West of Sgr A* (Fig.~\ref{f:pal_sou};
Fig.~\ref{f:mosaic}). In addition, the unusual red colors (K$_s$-[8.0])
of the three OB supergiants also indicate their associations with nearby molecular clouds
(\S~\ref{s:mid}). Such associations should be mostly real; the possibility for chance coincidence of the EMSs
with dense clouds is small, because the clouds have a small volume filling factor ($\sim 1\%$) 
in the GC~\citep{oka05}. The relatively small size ($\lesssim$25\arcsec,
i.e. $\lesssim$1 pc, see Fig.~\ref{f:mosaic}) and faintness of the associated
H{\small \rm II} regions further indicate that they are ionized by small groups
of massive stars or probably by the EMSs alone. The EMSs in H1, H2 and
H5 are spectroscopically identified as OB supergiants. The total extinction-corrected
Paschen-$\alpha$ luminosity of the H1
and H2 regions (see Fig.~\ref{f:mosaic}b) is
6.1$\times10^{36}~ergs~s^{-1}$, corresponding to a total
ionizing photon rate of $\sim4.1\times10^{49}~s^{-1}$~\citep{sco01}.
This rate can be accounted for by the two enclosed OB supergiants (P35 and P114; 
$Q_{Lyc}\sim10^{49.5}~s^{-1}$ each), consistent with the formation scenario that some of the
field PESs formed in small groups or in isolation. 

\section{Summary}\label{s:summary}

In this paper, we have investigated the properties of 180
PESs by using multi-wavelength
observations from near-IR to X-ray. These sources have been selected using their 
significant intensity enhancements in the F187N  narrow band, based on
our \hst /\nicm\ survey of the GC and previous \hst\
snapshot observations of the Arches and Central clusters. The multi-wavelength
data analysis has enabled us to probe the overall spectral characteristics
of individual sources, including a direct estimation (hence correction) 
of line-of-sight extinctions/absorptions. Our main results are as follows: 

\begin{itemize}
\item 10 sources are identified to be foreground stars. While a few of them
are apparently EMSs, the others with intrinsically low luminosities could be CVs.

\item For WNL stars, the Paschen-$\alpha$ line EW is well correlated with 
the intrinsic magnitude and color K$_s$-[3.6], which traces the free-free emission,
suggesting that the winds of the WNL are radiation-driven. 
However, the EW dependence on the magnitude appears to be sensitive to the stellar
age.

\item The above correlation seems to hold for WC stars in which the
  Paschen-$\alpha$ line EW is
greater than about 100 \AA. But those WC stars having lower EWs show
 anti-correlations with the magnitude and the K$_s$-[3.6] color, which can
be explained by an increased continuum level due to the strong hot dust 
emission from the stellar winds.

\item The presence of dusty stellar winds is clearly shown in the IR 
SEDs of a subset of the WC sample, showing intensity upturns or peaks in the 
mid-IR. From the
modeling of the SEDs, we estimate that the dust has a characteristic temperature of 
$\sim 750$ K and that each star has a dust mass in the range from $10^{-8}~M_{\odot}$ 
to a few $\times 10^{-5}~M_{\odot}$ (see \S~\ref{s:mid}). The upper end of this range is much higher than 
had previously been known for WC stars~\citep{vee98,mar99}, which may have important implications for understanding
the production of dust, especially in very young starburst galaxies at high
redshifts~\citep{dwe11}. 

\item We show that PESs may be roughly typed according to their positions
in the EW vs. magnitude diagram. This typing works `best' for WN stars. 
Both EW and magnitude ranges of WC stars are broad. Some WC 
stars have very low EWs, apparently due to enhanced continuum 
emission from hot circumstellar dust. They cannot be distinguished 
from OB supergiants based only on the EWs and magnitudes. We have 
tentatively assigned those unclassified PESs as dusty WC/OB
supergiants. Our detection of WR stars with relatively strong 
Pa$\alpha$ line emission should be quite complete. But it is 
possible that many Pa$\alpha$-faint WRs, especially WCs associated 
with substantial amounts of dust, remain to be detected.

\item Nearly half of the PESs are located outside the three clusters. The magnitude distribution of
these `field' sources shows two distinct peaks. One of the two peaks, containing roughly 
half of the sources, has a mean magnitude similar to those of the 
EMSs in the three clusters, while the other is substantially dimmer. 
The unclassified PESs in this latter peak have F190N magnitudes 
similar to or fainter than those of the WCs in the Central cluster. 
This suggests that the dimmer peak traces a stellar population 
with an average age similar to, or older than, that of the Central cluster.

\item Considering that the volume filling factor of dense clouds
  in the GC is very small,  we suggest that a few of the field PESs are physically
associated with compact H{\small \rm II} regions (with sizes 
$\lesssim$ 25\arcsec\ or 1 pc) and formed {\sl in situ},
 indicating the operation of a mode of massive star formation in 
small groups or even in isolation. 

\end{itemize}

\section*{Acknowledgments}
We are grateful to the referee, Andrea Stolte, for many valuable
comments, which helped greatly in the revision of the paper. We also
acknowledge the NASA support via the grant NNX11AI01G and the grant
HST-GO-11120 provided by the Space Telescope Science Institute, which
is operated by the Association of Universities for Research in
Astronomy, Inc., under NASA contract NAS 5-26555.


\begin{figure*}[!thb]
  \centerline{
        \epsfig{figure=fig1.ps,width=1\textwidth,angle=90}
       }
 \caption{Filter transmission curves of various instruments. Notice
that the filters of \hst /\nict\ are much 
   wider than those of \siri\ and \2mass, and that the J filter of \siri\ is
   narrower than that of \2mass. }
 \label{f:trans_curve}
 \end{figure*}

\begin{figure*}[!thb]
  \centerline{
       \epsfig{figure=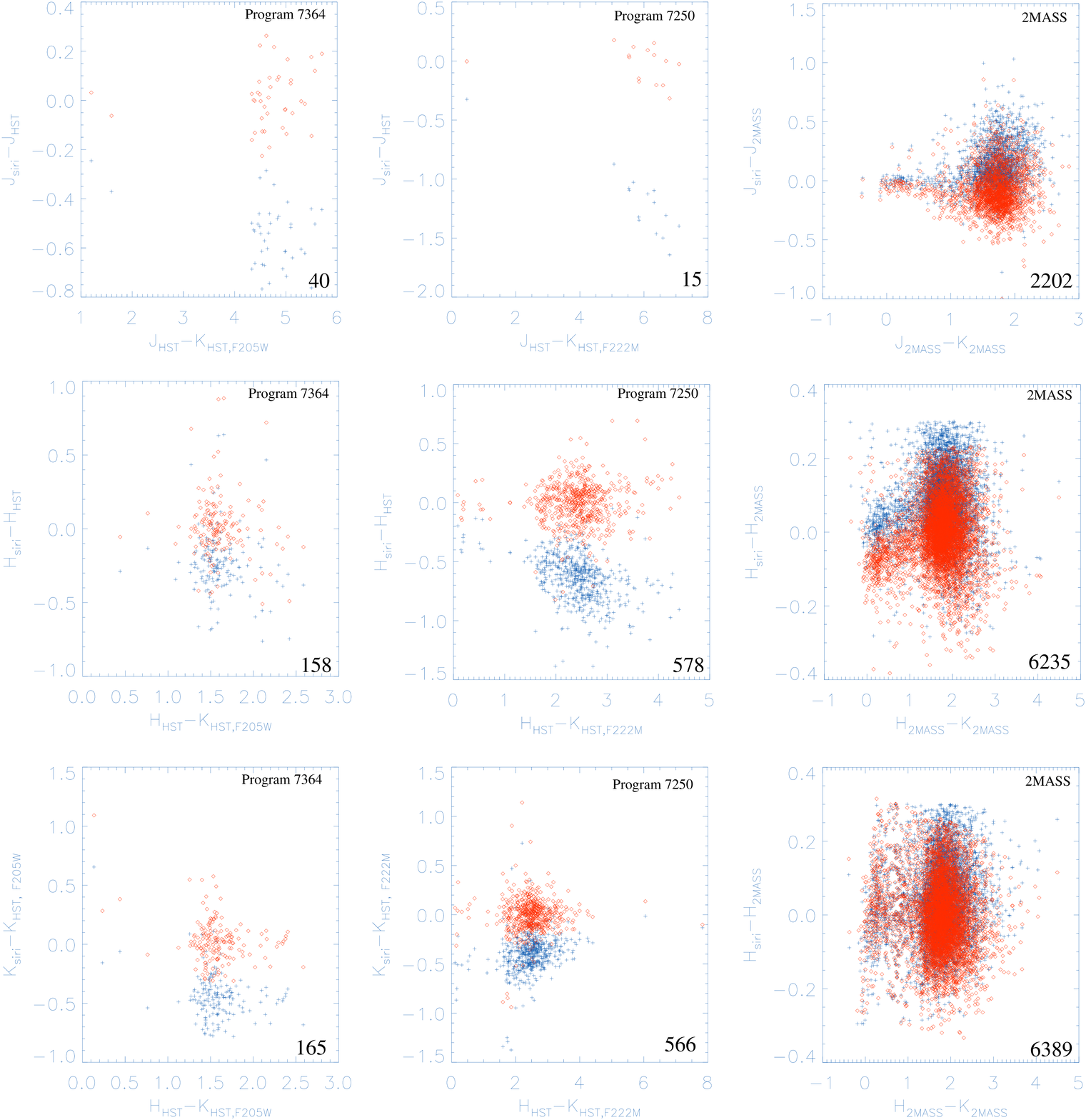,width=1\textwidth,angle=0}
       }
 \caption{The difference between the \siri\ and \hst /\nicm\ or 2MASS magnitudes as a function of the near-IR colors
   before (blue) and after (red) the magnitude transformation for the \hst /\nicm\
   magnitudes and 2MASS. The values in each lower right corners are the number of
 sources used to determine the magnitude transformations.}
 \label{f:mag_transfer}
 \end{figure*}

\begin{figure*}[!thb]
  \centerline{
       \epsfig{figure=fig3a.ps,width=0.5\textwidth,angle=90}
        \epsfig{figure=fig3b.ps,width=0.5\textwidth,angle=90}
      }
 \caption{(a) Comparison between the 1.90 $\mu$m extinctions of the PESs having
   available J, H and K$_s$ measurements, derived separately using J-H and H-K$_s$ colors. The black solid line 
represents A$_{F190N}$(J-H)=A$_{F190N}$(H-K$_s$). (b) Comparison
betweeh the A$_{F190N}$ values of the PESs with available J and H
magnitudes, derived separately using J-H and $\bar{r}$ (see
\S~\ref{ss:sample}. The ten potential foreground PESs identified in
\S~\ref{s:fore}, which have blue color: $J-H<2$, are excluded). The black solid line represents A$_{F190N}$(J-H)=A$_{F190N}$($\bar{r}$). }
\label{f:av_JHK_compare}
 \end{figure*}

\begin{figure*}[!thb]
  \centerline{
       \epsfig{figure=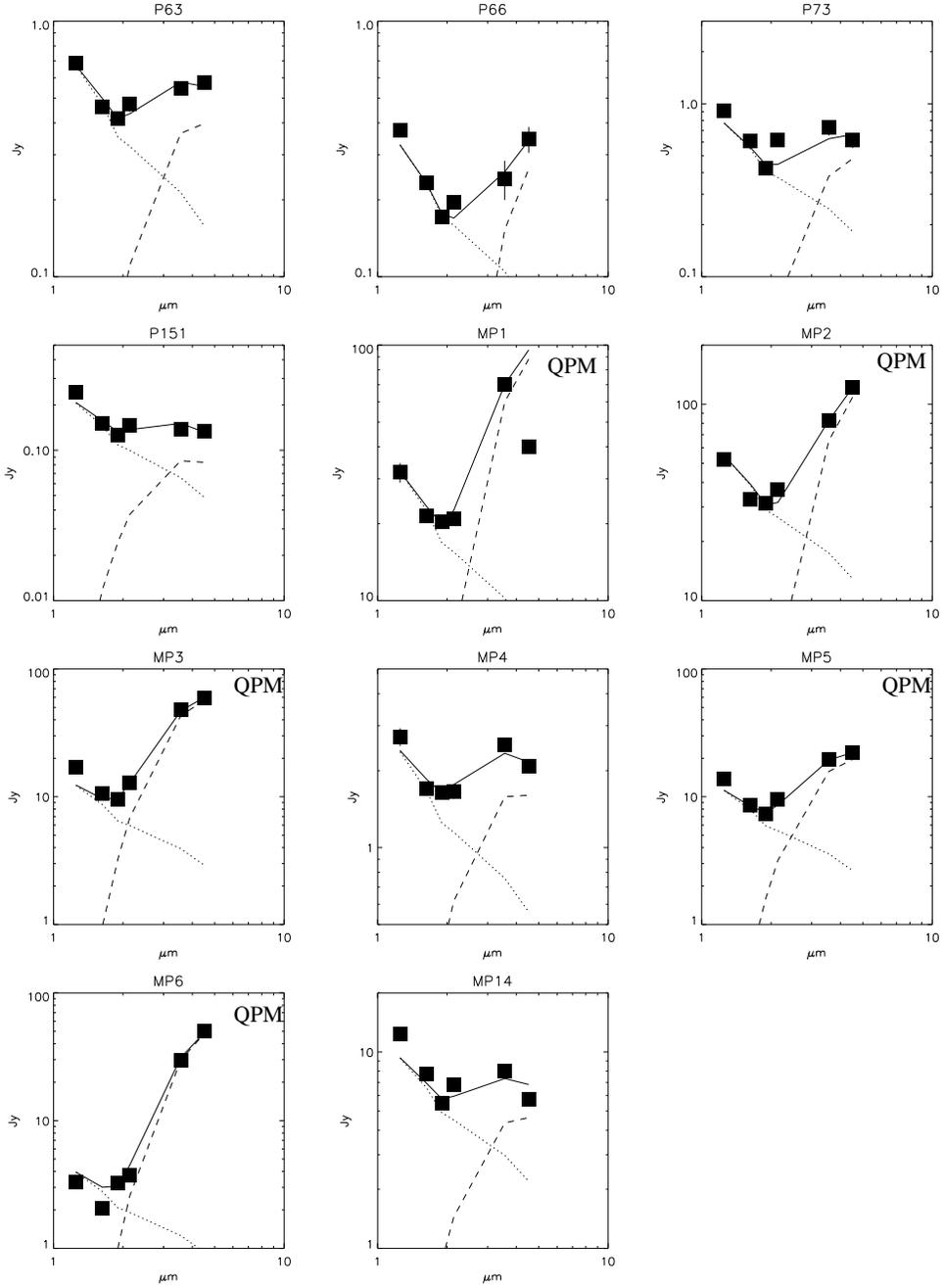,width=0.8\textwidth,angle=0}
       }
 \caption{The SED fitting result for 11 WC stars in six bands: J, H, F190N,
 $K_{s}$, 3.6 $\mu$m and 4.5 $\mu$m. The dotted lines represent the
 ``dust-free'' template of WC stars, while the 
dashed lines characterize the diluted dust blackbody 
contributions. The solid lines are the sum of the two
 components. QPMs have been flagged as such. All of them have strong
 mid-IR emission.}
 \label{f:SED_fitting}
 \end{figure*}

\begin{figure*}[!thb]
  \centerline{
       \epsfig{figure=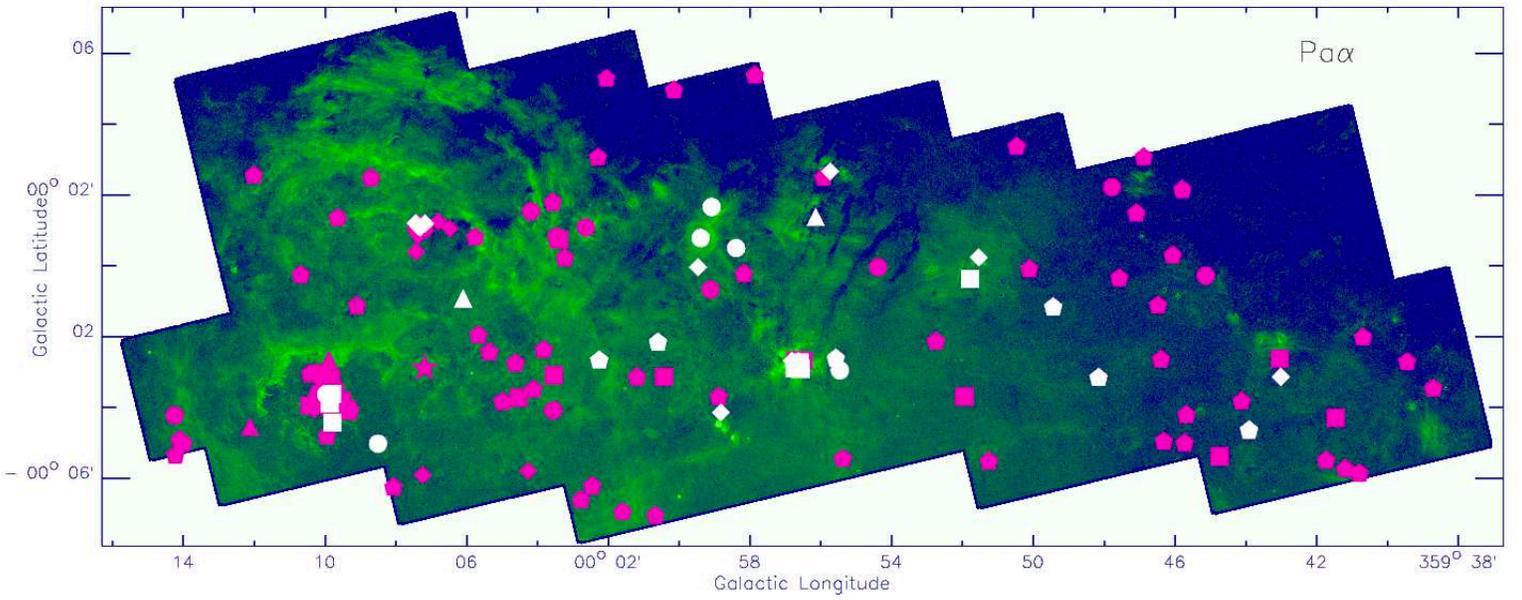,width=1\textwidth,angle=-90}
       }
 \caption{ 180 PESs overlaid on the mosaic image of Paschen-$\alpha$
   intensity (`diamond': WNL stars, `square': WC stars, `triangle':
   WNE stars, `circle':
   OB supergiants, `star symbols': LBV stars). The `pentagon' symbols are the stars without available
   spectroscopic identifications. The `white' symbols indicate the
   sources with X-ray counterparts (see \S~\ref{s:count_x}).}
 \label{f:pal_sou}
 \end{figure*}

\begin{figure*}[!thb]
  \centerline{
       \epsfig{figure=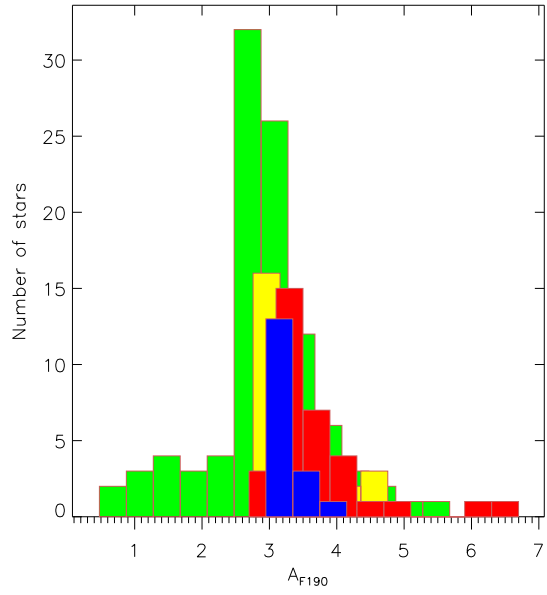,width=0.5\textwidth,angle=90}
       }
 \caption{The extinction distributions of the PESs in the Quintuplet
   (Yellow), Arches (Blue), Center (Red) and `field' (Green). 
PESs with $A_{F190N}<1.7$, i.e., H-K$_s$$<$1
   (see Eqn. 11) are considered to be foreground stars.}
\label{f:av_dis}
 \end{figure*}

\begin{figure*}[!thb]
  \centerline{
       \epsfig{figure=fig7a.ps,width=0.6\textwidth,angle=90}
       \epsfig{figure=fig7b.ps,width=0.6\textwidth,angle=90}
    }
 \caption{Left panel: Infrared color-color diagram of
   119 PESs with available J, H and K$_s$
   magnitude 
compared to the field stars 
from the \siri\ survey
 (small black dots). Right panel: Infrared magnitude-color diagram of
 45 PESs without J detections, compared to the field stars 
from the \siri\ survey
 (small black dots). Thick black lines delineate the region occupied by foreground
 stars (H-K $<$1 and J-H $<$ 2)}
 \label{f:IR_color}
 \end{figure*}

\begin{figure*}[!thb]
  \centerline{
       \epsfig{figure=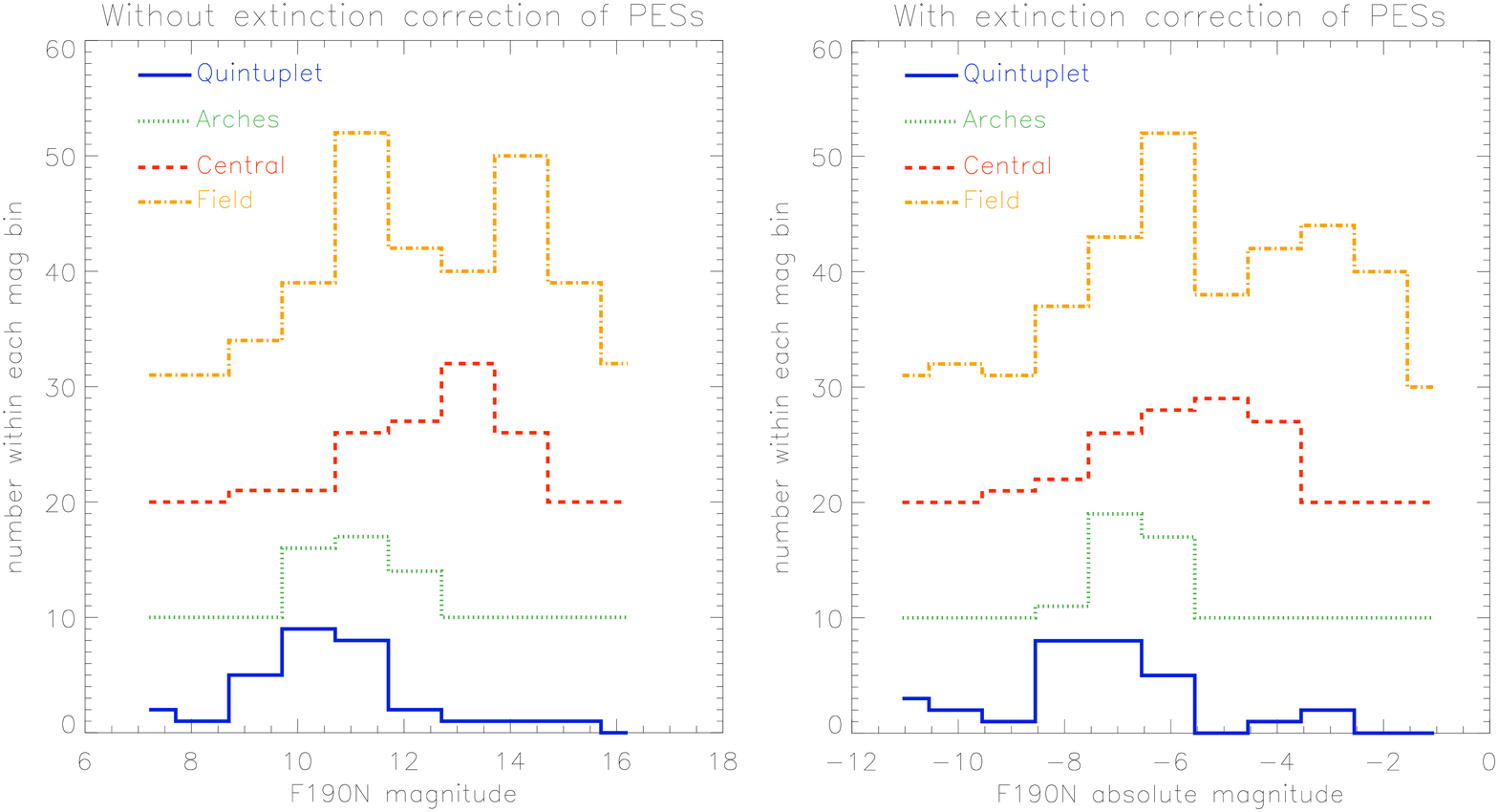,width=0.8\textwidth,angle=0}
       }
 \caption{F190 magnitude distributions before (left) and after (right)
the extinction and distance corrections (see \S~\ref{s:lum}). The
histograms for Arches, Central and `field' are shifted upward
by 10, 20 and 30 counts, respectively.}
 \label{f:lum_pal}
 \end{figure*}

\begin{figure*}[!thb]
  \centerline{
       \epsfig{figure=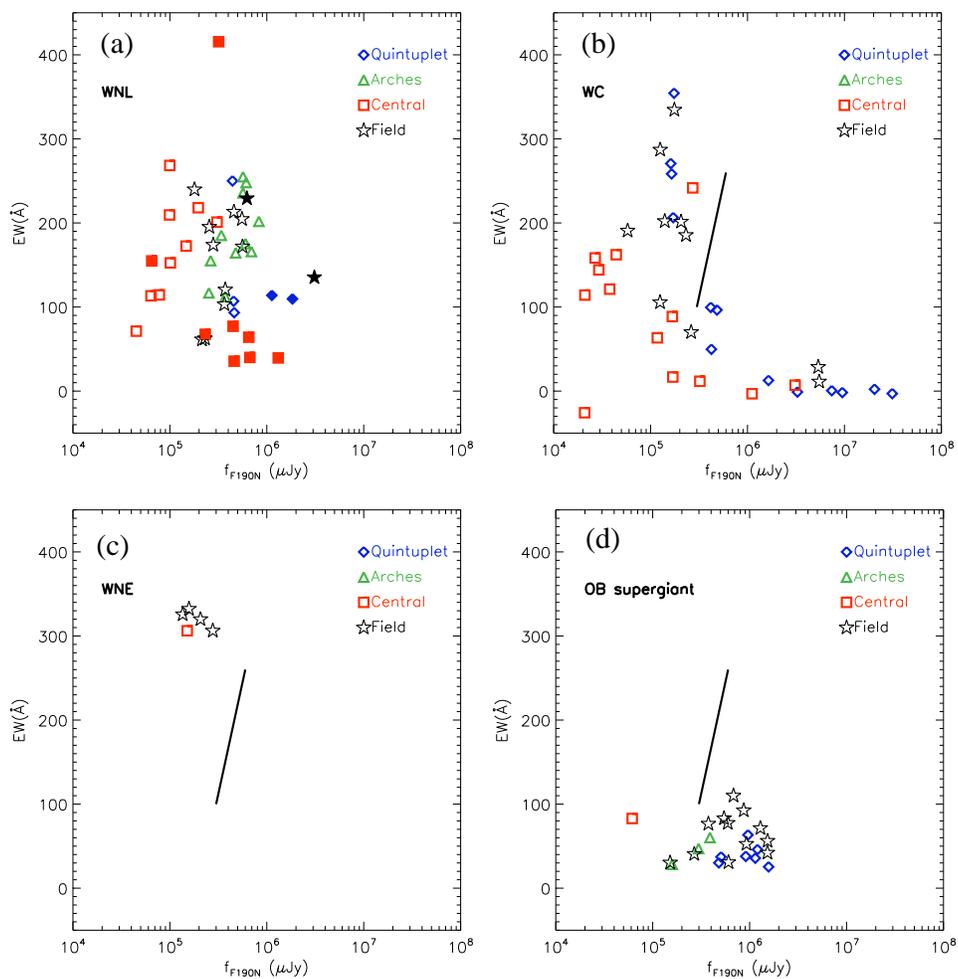,width=0.8\textwidth,angle=0}
       }
 \caption{Paschen-$\alpha$ equivalent width (EW)
   vs. extinction-corrected F190N flux densities, $f^o_{F190N}$, plots of stars of type WNL (a), WC (b), WNE (c) and OB supergiant (d). In (a), the filled symbols
are the Ofpe/WN9 stars. The solid lines in (b)-(d) represent the positive correlation 
between the EW and $f_{F190N}$, seen in (a) for the WNL stars in the Arches cluster.}
 \label{f:ew}
 \end{figure*}

\begin{figure*}[!thb]
  \centerline{
       \epsfig{figure=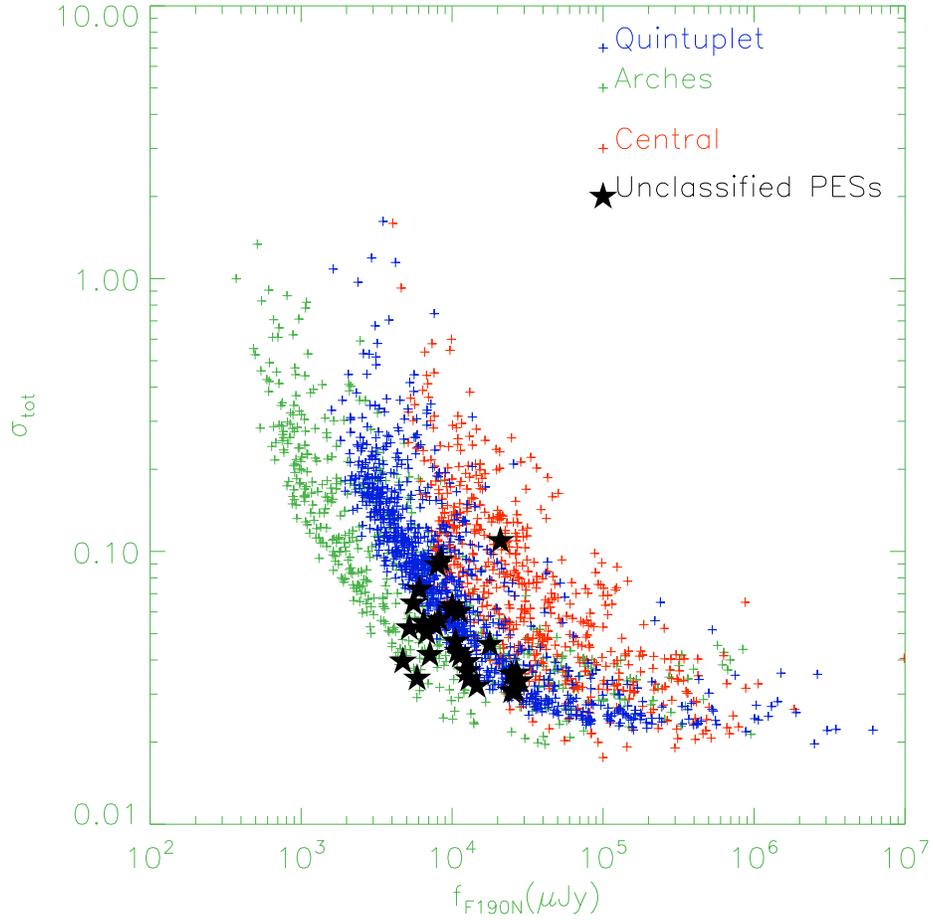,width=0.8\textwidth,angle=90}
       }
 \caption{Comparison of the uncertainty of the flux ratio
   (r=$\frac{f_{F187N}}{f_{F190N}}$), $\sigma_{tot}$, (see
   \S~\ref{ss:sample}) 
between the normal sources
   within the three clusters and the unclassified
   PESs, which have low EW and $f^o_{F190N}$. }
 \label{f:f_uf}
\end{figure*}

\begin{figure*}[!thb]
  \centerline{
       \epsfig{figure=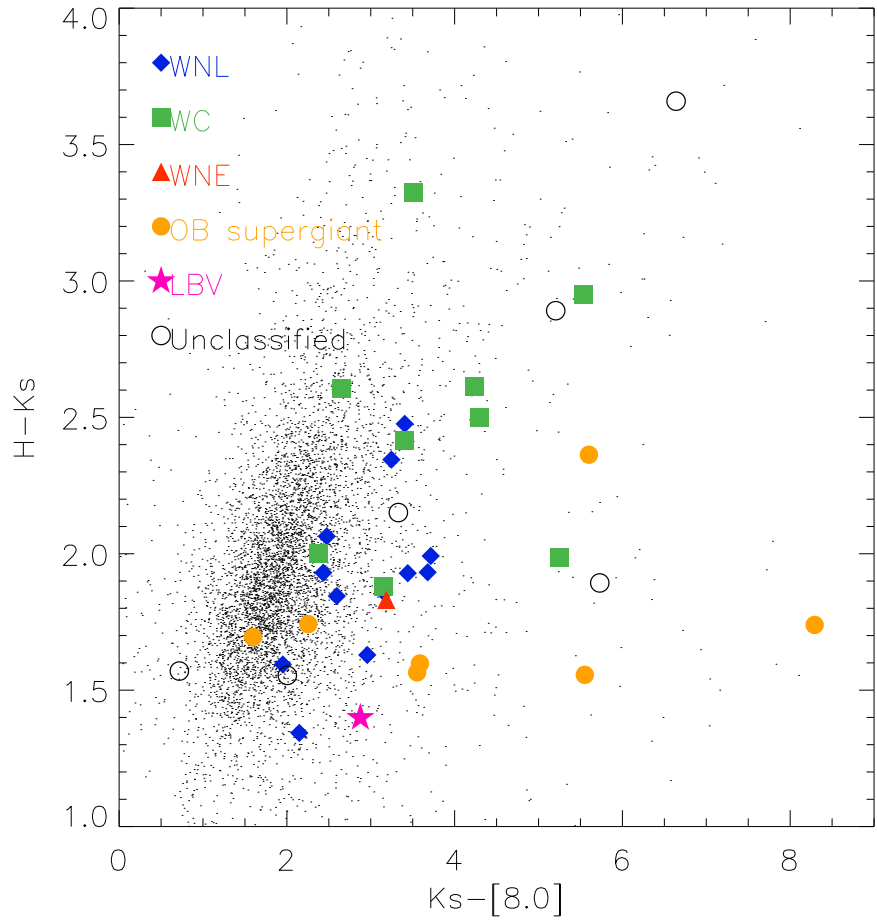,width=0.8\textwidth,angle=90}
       }
 \caption{H-K$_s$ vs. K$_s$-[8.0] plot for 36
   PESs and field stars from the \spit/\irac\ GALCEN survey (black dots).}
 \label{f:Mid_color}
 \end{figure*}

\begin{figure*}[!thb]
  \centerline{
       \epsfig{figure=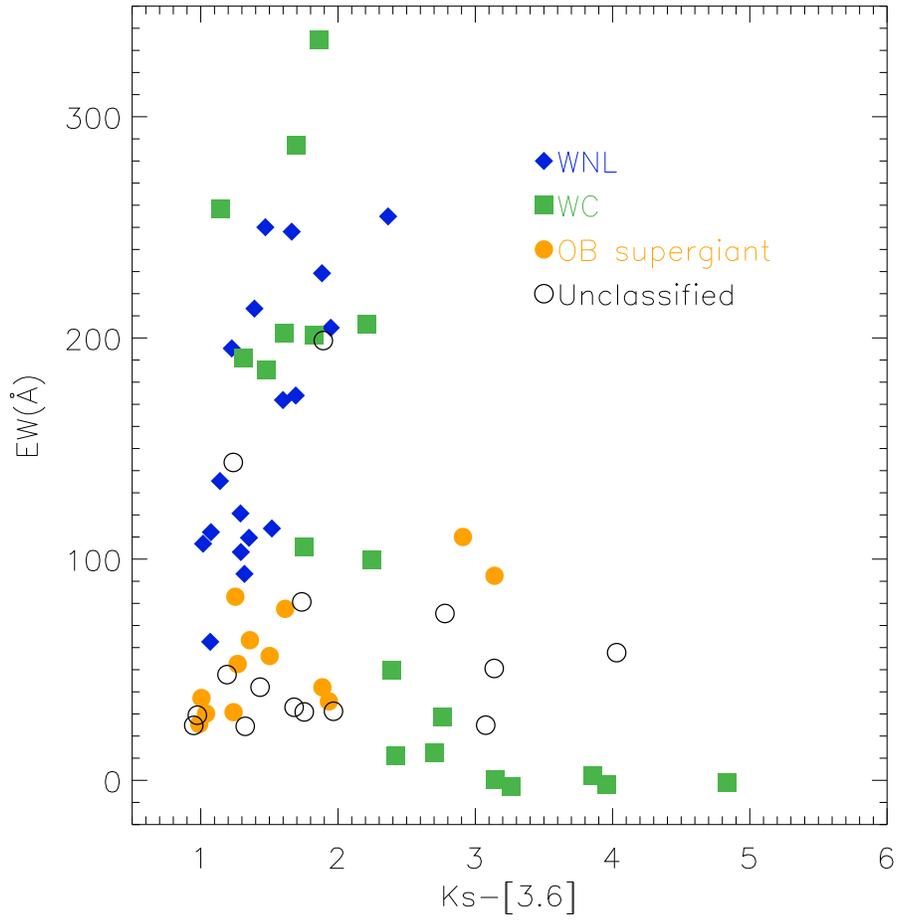,width=0.8\textwidth,angle=90}
       }
 \caption{ EW vs. K$_s$-[3.6] plot for various types of PESs.}
 \label{f:Mid_color_1}
 \end{figure*}

\begin{figure*}[!thb]
  \centerline{
       \epsfig{figure=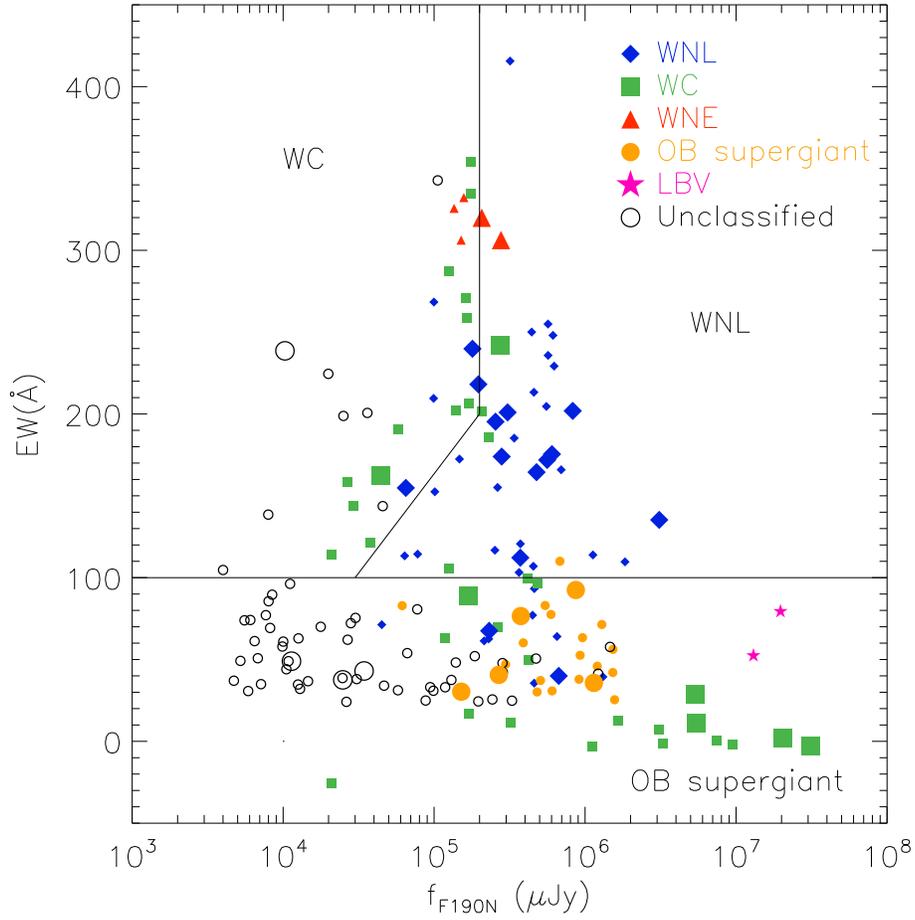,width=0.8\textwidth,angle=90}
       }
 \caption{The combined Paschen-$\alpha$ EW vs. $f^o_{F190N}$ plot 
of stars of various types. Lines are drawn to roughly show the regions dominated by 
WNL, WC and OB supergiants. The stars with larger symbols are the ones with X-ray counterparts.
          }
 \label{f:pal_all}
 \end{figure*}

\begin{figure*}[!thb]
  \centerline{
       \epsfig{figure=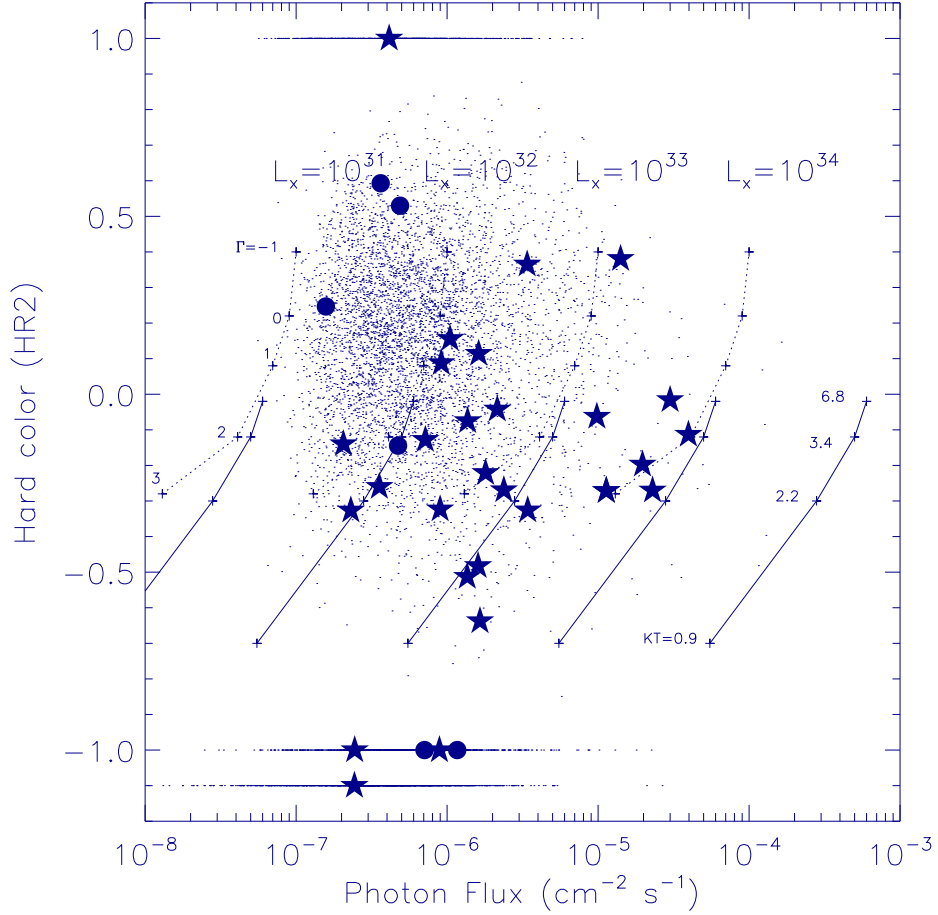,width=0.8\textwidth,angle=90}
       }
 \caption{HR2 vs. Photon flux plot for the 34 X-ray counterparts of the
 PESs. The blue dots are from the \chan\ X-ray
 Catalog~\citep{mun09}. Most of them 
have high HR2 ($>$0) and low photon flux ($<10^{-6}~cm^{-2}~s^{-1}$)
and should be CVs. The
 `star' symbol represents the counterparts that have been reported in
 the 
literature, while the `circle' symbols are the new X-ray
counterparts. We also label the positions for different types of spectra,
either power-law with index $\Gamma$ or thermal plasma at temperature T, as in~\citet{mau10c}}
 \label{f:X_color}
 \end{figure*}

\begin{figure*}[!thb]
  \centerline{
       \epsfig{figure=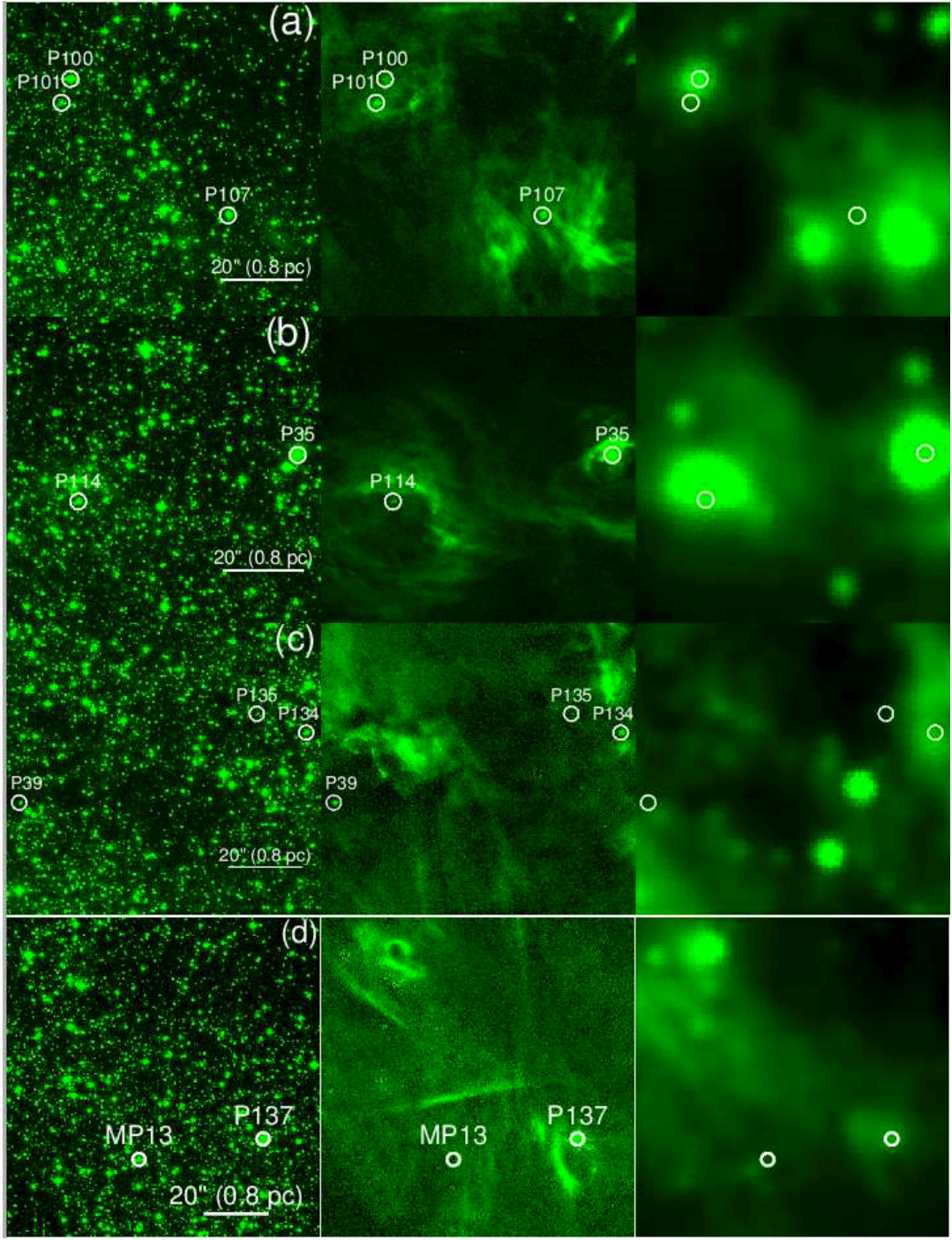,width=0.9\textwidth,angle=0}
       }
 \caption{\hst\ and Spitzer images of four H{\small \rm II} regions in F190N
(left panel), Paschen-$\alpha$ (middle) and 8.0
   $\mu$m (right). From top to bottom: (a) H5, (b) H1 and H2, (c)
   (l,b)=(-0.06,0.02), (d) (l,b)=(-0.13,0.0)}
 \label{f:mosaic}
 \end{figure*}

\begin{center}
\begin{longtable}{|c|c|}
\caption{Acronym/Abbreviation List}\label{t:acronym}\\

\endhead 

\hline
CV & Cataclysmic variable \\
EMS & Evolved massive star \\
EW & Equivalent width \\
FWHM & Full width at half maximum \\
GC & Galactic center \\
HST & Hubble Space Telescope \\
IRAC & Infrared array camera \\
ISM & Interstellar medium \\
LBV & Luminous blue variable \\
MS & Main sequence \\
NICMOS & Near infrared camera and multi-object spectrometer \\
PAH & Polycyclic aromatic hydrocarbon \\
PES & Paschen-$\alpha$ emitting source \\
QPM & Quintuplet-proper member \\ 
SED & Spectral energy distribution \\
SN & Supernova \\
WR & Wolf-Rayet \\
WN & Nitrogen sequence Wolf-Rayet \\
WNE & Early type Nitrogen sequence Wolf-Rayet \\
WNL & Late type Nitrogen sequence Wolf-Rayet \\
WC &  Carbon sequence Wolf-Rayet\\
WCE & Early type Carbon sequence Wolf-Rayet\\
WCL & Late type Carbon sequence Wolf-Rayet\\
\hline
\end{longtable}
\end{center}

\tabcolsep=0.09cm
\begin{deluxetable}{cccccccccccccccccc}%
  \tabletypesize{\tiny}
 \rotate
 \tablecolumns{18}
  \tablecaption{Catalog of Paschen-$\alpha$ emission-line stars}
  \tablewidth{0pt}
  \tablehead{
  \colhead{Name} & 
  \colhead{R.A.} &
  \colhead{Dec.} &
  \multicolumn{9}{c}{Magnitude} &
  \colhead{$A$} &
  \colhead{EW} &
  \colhead{$N_s$} & 
  \colhead{Counterpart} &
  \colhead{Type} &
  \colhead{Location}  \\
  \colhead{} & 
  \colhead{(J2000)} &
  \colhead{(J2000)} &
  \colhead{J} &
  \colhead{H} &
  \colhead{K} & 
  \colhead{[1.87]} &
  \colhead{[1.9]} &
 \colhead{[3.6]}& 
 \colhead{[4.5]}& 
 \colhead{[5.4]}& 
 \colhead{[8.0]}&
  \colhead{} &
  \colhead{} &
  \colhead{} & 
  \colhead{} &
  \colhead{} &
  \colhead{}\\
  }
  \startdata
   P1&266.62478&-28.78001&          13.4(0.02)&          12.8(0.01)&          12.6(0.02)&          12.5(0.02)&          12.6(0.02)&                    &                    &                    &                    &     0.5&      17.0(5.0)& 5.6&                    &                    &F\\
   P2&266.59921&-28.80299&          16.3(0.05)&          13.3(0.01)&          11.4(0.02)&          11.1(0.02)&          12.1(0.02)&           9.9(0.06)&           9.1(0.04)&           8.9(0.03)&           8.2(0.05)&     2.8&    332.2(12.4)&26.7&Mau10c\_17          &   WN5b             &F\\
   P3&266.51501&-28.78606&          17.5(0.11)&          14.1(0.04)&          12.3(0.03)&          12.9(0.02)&          13.0(0.02)&          10.5(0.07)&          10.5(0.07)&          10.1(0.05)&                    &     3.2&      30.9(5.4)& 4.9&                    &                    &F\\
   P4&266.55606&-28.81641&          14.6(0.02)&          11.5(0.01)&           9.8(0.02)&          10.2(0.02)&          10.5(0.02)&           8.4(0.04)&           8.0(0.04)&           7.9(0.02)&                    &     3.2&      63.3(6.0)&10.3&FQ\_381             &   OBI              &Q\\
   P5&266.55408&-28.82014&      17.5(0.07)$^h$&      14.4(0.06)$^h$&      12.5(0.06)$^h$&          13.6(0.02)&          13.8(0.02)&                    &                    &                    &                    &     3.2&      34.0(5.7)& 5.9&                    &                    &Q\\
   P6&266.56301&-28.82693&      13.5(0.06)$^h$&      10.4(0.05)$^h$&       8.6(0.06)$^h$&           9.1(0.02)&           9.6(0.02)&           7.3(0.09)&           6.8(0.11)&           7.0(0.06)&                    &     2.9&     109.7(7.1)&15.1&Lie\_71,FQ\_241     &   WN9/Ofpe         &Q\\
\enddata
\label{t:catalog}
\tablecomments{R.A. and Decl. are in degrees. J, H and K band magnitudes are mainly from the SIRIUS catalog~\citep{nis06}, 
except for those with the superscript `h' and `t', which are from the \hst /\nics\ snapshot observations and 2MASS catalog (\S~\ref{s:count_ir}). The extinction $A$ values at 1.90$\mu$m are from the near-IR colors, except for 
those with the superscript `b', which are from the ratio map in~\citet{don11} (\S~\ref{s:extinction}). 
EW is in units of \AA.
The spectroscopically identified counterpart names and types are from ~\citet{fig99a} (FQ\_*),~\citet{fig02}(FA\_*),~\citet{lie09}(Lie\_*),~\citet{pau06}(E*),~\citet{blu01}(Blu01\_*),~\citet{cot99}(Cot\_*),~\citet{hom03}(Ho*),~\citet{mun06}(Muno06\_*),~\citet{mik06}(Mik06\_*),~\citet{mau07}(Mau07\_*),~\citet{mau10a}(Mau10a\_*),~\citet{mau10b}(Mau10b\_*),~\citet{mau10c}(Mau10c\_*). If 
multiple identifications are available, we adopt the types from the latest reference, except for P6 and P7, which are from ~\citet{fig99a}, instead of~\citet{lie09}. The `Location' column divides the sources into four groups: those inside the three clusters (`Q' - Quintuplet, `A'- Arches, `C' - Center) and outside  (`F'; \S~\ref{s:observation}). The MP1-14 are not selected out by our method ($N_s~<$ 4.5, see \S~\ref{ss:sample}). Instead, they are WC stars identified by previous ground-based spectroscopic observations. We include these WC stars into our catalog to complete the sample of the evolved massive stars in the GC}
 \end{deluxetable}

\begin{deluxetable}{cccc}
 \tabletypesize{\footnotesize}
  \tablecaption{Previous HST NIC2 observations}
  \tablewidth{0pt}
  \tablehead{
  \colhead{Program ID} &
  \colhead{Observation Data} &
  \colhead{Filter} &
  \colhead{exposure} \\
 }
\startdata
7250 & Aug 31st, 1999& F187N & 1215 sec \\
7250 & Aug 31st, 1999& F190N & 1215 sec \\
7364 & Sep 14th, 1999& F187N & 1024 sec \\
7364 & Sep 14th, 1999& F190N & 1024 sec \\ 
\enddata
\label{t:pre_obs}
\end{deluxetable}

\begin{deluxetable}{ccccccc}
  \tabletypesize{\tiny}
  \tablecaption{The intrinsic near-IR colors of massive stars}
  \tablewidth{0pt}
  \tablehead{
  \colhead{Type} & 
 \colhead{Description} & 
 \colhead{$(J-H)_{0}$} &
 \colhead{$(J-H)_{0}$ Error} &
  \colhead{$(H-K_s)_{0}$}  &
 \colhead{$(H-K_s)_{0}$ Error}  \\
  }
  \startdata
  WN5-6 weak & WN5, WN6 & 0.03 & 0.012 & 0.16 & 0.012\\
  WN4-7 strong & WN4b, WN5b, WN6b & 0.10 & 0.006 & 0.27 & 0.006\\
  WN7-9 weak & WN7(h), WN8(h), WN9(h) & 0.02 & 0.052 & 0.11 & 0.053\\
  WN10-11 & Ofpe/WN9, WN10, WN11 & 0.06 & 0.049 & 0.13 & 0.11 \\
  WC5-7 & WC5, WC6, WC7 & 0.04 & 0.029 & 0.58 & 0.012\\
  WC8 & WC8 & 0.05 & 0 & 0.38 & 0\\
  WC9 & WC9 & -0.03 & 0 & 0.26 & 0\\
  LBV & LBV & -0.07 & 0 & -0.05 & 0\\
  OB supergiant & O If+, B[e] & -0.11 & 0 & -0.1 & 0\\ 
 \enddata
\label{t:color}
\tablecomments{The colors of the WNs and WCs are from~\citet{cro06}. According to~\citet{fig98}, we assume the colors of LBV are similar to O9.5I~\citep{koo83}.~\citet{mar06} provides the colors of OB supergiant}
 \end{deluxetable}
\begin{deluxetable}{ccccccccc}
  \tabletypesize{\tiny}
  \tablecaption{X-ray counterparts of the Paschen-$\alpha$ emitting sources}
  \tablewidth{0pt}
  \tablehead{
  \colhead{Name} & 
  \colhead{Reference} &
  \colhead{X-ray ID} &
  \colhead{$\sigma_x$ (arcsec)} &
  \colhead{X-ray flux} &
  \colhead{HR0} &
  \colhead{HR1} &
  \colhead{HR2} &
  \colhead{P (\%)} \\
  }
  \startdata
   P9&Lie\_96,Mau10a\_19  &CXOGC174615.1-284932&   0.51 & 1.4e-06&$  1.00^{+0.23}_{-9.00}$&$ -0.16^{+0.28}_{-0.26}$&$ -0.51^{+0.36}_{-0.35}$ & 43.5\\
  P18&    FA\_2,Mau10a\_11&CXOGC174549.7-284925&   0.41 & 1.6e-06&$  1.00^{+0.66}_{-9.00}$&$  0.56^{+0.36}_{-0.39}$&$  0.11^{+0.32}_{-0.31}$ & 31.8 \\
  P35&Cot\_4,Mau10a\_7    &CXOGC174528.6-285605&   0.48& 9.0e-07&$  0.82^{+0.13}_{-0.12}$&$ -0.12^{+0.15}_{-0.14}$&$ -0.32^{+0.18}_{-0.18}$ & 41.4\\
  P36&Mau10c\_6           &CXOGC174531.4-285716&   0.50 & 2.4e-07&$  0.57^{+0.31}_{-0.28}$&$ -0.53^{+0.44}_{-0.29}$&$ -1.00^{-9.00}_{-1.15}$ & 44.8\\
  P38&Mau10c\_7           &CXOGC174537.9-290134&   0.33& 1.4e-06&$ -0.17^{+0.10}_{-0.11}$&$ -0.57^{+0.13}_{-0.13}$&$ -0.07^{+0.26}_{-0.26}$ & N/A\\
  P39&Mau10a\_6           &CXOGC174522.6-285844&  0.63 &  2.3e-07&$  1.00^{+0.58}_{-9.00}$&$  0.31^{+0.32}_{-0.57}$&$ -0.33^{+0.44}_{-0.34}$ & 21.4\\
  P75&M07\_2,Mau10a\_22   &CXOGC174617.0-285131&  0.56 &  2.4e-06&$  1.00^{+0.22}_{-9.00}$&$ -0.05^{+0.22}_{-0.21}$&$ -0.27^{+0.26}_{-0.25}$ & 31.9\\
  P80&    FA\_6,Mau10a\_13&CXOGC174550.4-284922&   0.33 & 4.0e-05&$  0.77^{+0.06}_{-0.05}$&$  0.02^{+0.06}_{-0.06}$&$ -0.11^{+0.07}_{-0.07}$ & 84.0\\
\enddata
\label{t:pal_x}
\tablecomments{The X-ray ID, flux (photons $s^{-1}$ $cm^{-2}$, 0.5-8 keV) and X-ray colors (HR) are from \citet{mun09}. HR=$\frac{h-s}{h+s}$. For HR0, HR1 and HR2, h (s) represents the X-ray photons in the 2.0-3.3 keV (0.5-2.0 keV), 3.3-4.7 keV (2.0-3.3 keV), 4.7-8.0 keV (3.3-4.7 keV). `1' (`-1') means that the source is not been detected in the soft (high) band. '-9' means that the source is not detected in both h and s bands. $\sigma_x$ is the X-ray source positional uncertainty. The `P (\%)' column lists the percent probability for the infrared object as real counterpart of the X-ray corresponding source as listed in Table 3 of~\citet{mau09}. 'N/A' means that the PES is outside their adopted match radii of the X-ray sources. `-1' means that the PES has a red near-IR color (H-K$_s$$>$0.9), while the X-ray source is soft  (HR0$<$-0.175). We find that P145 actually has a blue near-IR color (H-K$_s$=0.8), based on 2MASS measurements, although a red color was adopted in~\citet{mau09}, using the SIRIUS catalog, in which the J and H band fluxes are saturated.}  
 \end{deluxetable}
\begin{deluxetable}{ccccc}
  \tabletypesize{\tiny}
  \tablecaption{Modified $N_s$ for the foreground stars}
  \tablewidth{0pt}
  \tablehead{
  \colhead{Name} & 
  \colhead{Counterpart} &
  \colhead{Extinction-free $m_{F190N}$} &
  \colhead{$A_{F190N}$} &
  \colhead{Revised $N_s$} \\
  }
  \startdata
\multicolumn{5}{c}{First Class} \\ \hline
  P38&Mau10c\_7           & 7.6& 1.5& 2.9\\
 P140&Mau10c\_1           & 5.9& 1.3& 9.8\\
 P145&                    & 7.0& 1.3& 3.3\\
\multicolumn{5}{c}{Second Class} \\ \hline
   P1&                    &12.2& 0.5& 3.4\\
  P27&                    &13.8& 1.5& 7.0\\
 P58&                    &10.2& 1.6& 4.1\\
 P102&                    &11.2& 1.6& 7.2\\
 P116&                    &12.4& 0.5& 2.7\\
 P149&                    &11.8& 1.1&13.6\\
 \enddata
\tablecomments{$N_s$ is the parameter which we use to identify the
  PESs in \S~\ref{ss:sample}. In~\citet{don11}, we only considered the
sources with $N_s>4.5$ as PESs. We divide the nine sources with blue
near-IR colors into two class according to their extinction-corrected
free $m_{F190N}$ (see \S~\ref{s:fore}). The first class should be foreground massive stars,
while the second class could be foreground CVs instead.}
\label{t:fore}
 \end{deluxetable}
\begin{deluxetable}{ccccccc}
  \tabletypesize{\small}
  \tablecolumns{7}
  \tablecaption{Dust properties in the WC stars}
  \tablewidth{0pt}
  \tablehead{
  \colhead{Name} & 
  \colhead{T$_{dust}$ (K)} & 
  \colhead{F190N dust percentage} &
  \colhead{Unc EW} &
  \colhead{Corr EW} &
  \colhead{Dust Mass ($M_{\odot}$)} \\
  }
  \startdata
 P63 & 834$\pm$53& 15.5$\pm$  4.7& 99.5$\pm$6.9&120.5$\pm$  11.4&   3.5 $\times10^{-8}$\\
 P66 & 574$\pm$91&  2.2$\pm$  2.2&206.4$\pm$13.4&211.9$\pm$  14.8&   1.2 $\times10^{-7}$\\
 P73 & 727$\pm$44&  8.1$\pm$  2.9& 49.7$\pm$8.1& 56.1$\pm$  9.5&   6.6 $\times10^{-8}$\\
 P151& 944$\pm$25& 18.4$\pm$  1.8&105.6$\pm$10.0&134.3$\pm$  13.5&   5.1 $\times10^{-9}$\\
 MP1 & 644$\pm$28& 15.3$\pm$  3.5&  2.2$\pm$4.6&  6.4$\pm$  5.5&   1.9 $\times10^{-5}$\\
 MP2 & 586$\pm$ 6&  6.2$\pm$  0.5& -3.0$\pm$4.4& -1.6$\pm$  4.7&   3.8 $\times10^{-5}$\\
 MP3 & 694$\pm$ 8& 33.4$\pm$  2.5& -1.7$\pm$4.5&  7.6$\pm$  7.0&   8.0 $\times10^{-6}$\\
 MP4 & 905$\pm$39& 23.7$\pm$  4.8& 12.7$\pm$4.8& 20.9$\pm$  6.7&   9.3 $\times10^{-8}$\\
 MP5 & 739$\pm$15& 21.5$\pm$  2.6&  0.5$\pm$4.5&  5.8$\pm$  5.9&   2.2 $\times10^{-6}$\\
 MP6 & 599$\pm$ 7& 32.6$\pm$  2.6& -1.1$\pm$4.5&  9.0$\pm$  6.8&   1.4 $\times10^{-5}$\\
 MP14& 858$\pm$25& 15.0$\pm$  2.2& 11.2$\pm$4.8& 16.1$\pm$  5.8&   3.1 $\times10^{-7}$\\
 \enddata
\tablecomments{Dust properties derived from the SED fitting of 11 WCs
  in six bands: J, H, F190N, $K_s$, 3.6$\mu$m and 4.5 $\mu$m. The second
  column presents the single dust temperature, T$_{dust}$ (K). The 
percentage of the total F190N flux attributable to dust is listed in the third
  column. The fourth column shows the uncorrected EW as listed in
  Table~\ref{t:catalog}, while the fifth column presents the corrected
  value (after subtracting the dust contribution). The derived dust mass is listed in the sixth column.}
\label{t:dust}
 \end{deluxetable}

\end{document}